\documentclass[fleqn,10pt]{wlscirep}
\usepackage{subcaption}
\usepackage{multirow}
\usepackage{xcolor}
\usepackage{longtable}
\usepackage{siunitx}

\title{Using Machine Learning to Predict the Evolution of Physics Research}
\author[1,2*,+]{Wenyuan Liu}
\author[3**,+]{Stanisław Saganowski}
\author[3]{Przemysław Kazienko}
\author[1,2]{Siew Ann Cheong}
\affil[1]{School of Physical and Mathematical Sciences, Nanyang Technological University, 21 Nanyang Link, Singapore 637371, Republic of Singapore}
\affil[2]{Complexity Institute, Nanyang Technological University, 61 Nanyang Drive, Singapore 637335, Republic of Singapore}
\affil[3]{Department of Computational Intelligence, Faculty of Computer Science and Management, Wrocław University of Science and Technology, Wrocław, Poland}
\affil[*]{liuw0037@ntu.edu.sg}
\affil[**]{stanislaw.saganowski@pwr.edu.pl}
\affil[+]{these authors contributed equally to this work}

\begin{abstract}
The advancement of science as outlined by Popper and Kuhn is largely qualitative, but with bibliometric data it is possible and desirable to develop a quantitative picture of scientific progress.
Furthermore it is also important to allocate finite resources to research topics that have growth potential, to accelerate the process from scientific breakthroughs to technological innovations.
In this paper, we address this problem of quantitative knowledge evolution by analysing the APS publication data set from 1981 to 2010.
We build the bibliographic coupling and co-citation networks, use the Louvain method to detect topical clusters (TCs) in each year, measure the similarity of TCs in consecutive years, and visualize the results as alluvial diagrams.
Having the predictive features describing a given TC and its known evolution in the next year, we can train a machine learning model to predict future changes of TCs, i.e., their continuing, dissolving, merging and splitting.
We found the number of papers from certain journals, the degree, closeness, and betweenness to be the most predictive features.
Additionally, betweenness increases significantly for merging events, and decreases significantly for splitting events.
Our results represent a first step from a descriptive understanding of the Science of Science (SciSci), towards one that is ultimately prescriptive.
\end{abstract}

\begin{document}

\flushbottom
\maketitle

\thispagestyle{empty}

\section*{Introduction}
We all become scientists because we want to create an impact and make a difference, to the lives of those around us, and also to the many generations that are to come.
We all strive to make choices in the problems we study, but not all choices lead to breakthroughs.
There is actually a lot more about scientific breakthroughs that we can try to understand.
For one, science is an ecosystem of scholars, ideas, and papers published.
In this ecosystem, scientists can form strongly-interacting groups over a particular period to solve specific problems, but later drift apart as their interests diverge, or due to the availability or paucity of funds, or other factors.
The evolution of these problem-driven groups is more or less completely documented by the papers published as outcomes of their research.
By analysing groups of closely related papers, researchers could extract rich information about knowledge processes \cite{Chen2010, Rosvall2010, Liu2017a}.
The potential to map scientific progress using publication data has attracted enormous interest recently \cite{Zeng2017, Fortunato2018, Hicks2015}.
However, compared to the study on science at the level of individual papers \cite{Radicchi2008, Wang2013, Ke2015} and at the level of the whole citation network \cite{Small1999, Boyack2005, Bollen2009}, where a lot of work has already been done, the research on science at the community level is still limited \cite{Chen2010, Liu2017a}.

In a recent paper, Liu \emph{et al.} demonstrated the utility of visualizing and analysing scientific knowledge evolution for physics at the aggregated mesoscale through the use of alluvial diagrams\cite{Liu2017a}.
In this picture, papers are clustered into groups (or communities) and these groups can grow or shrink, merge or split, new groups may arise while the others may dissolve.
This shares a very strong parallel with what some researchers discovered in social group dynamics \cite{Palla2007}.
More importantly, many breakthroughs were made by scientists absorbing knowledge from other fields, often in a very short time.
On the alluvial diagrams, these knowledge transformations manifest themselves as merging and splitting events.
Clearly, funding agencies, universities and research institutes would want to promote growing research fields, and particularly those where breakthroughs are imminent.
This is why it is important to be able to predict the future events.
Liu \emph{et al.}\cite{Liu2017a} attempted this in their paper, by analysing the correlation between event types and several network metrics.
Unfortunately, such predictions are very noisy.
While merging events are highly correlated with interconnections between communities, the correlation between splitting events and the internal structure of communities are much more complex; besides, the predictions of forming, dissolving, growing, shrinking were not considered at all.

Given the recent successes in the area of machine learning and artificial intelligence to a variety of prediction problems \cite{Carrasquilla2017, Ahneman2018}, as well as having developed and validated a general framework to predict social group evolution in Saganowski \emph{et al.}\cite{Saganowski2017}, we decided to utilize machine learning techniques to fill the gap in predicting scientific knowledge events \cite{Saganowski2015, Ilhan2016, Pavlopoulou2017}.
The overall idea behind the Group Evolution Prediction (GEP) method is to build a classification model trained with historical observations in order to predict the future group changes based on their current characteristics, such as size, density, average degree of nodes, etc.
A single historical observation consists of a set of features describing the group at a given point in time, and an event type that this group just experienced.
The profile of the group may reflect its structure (e.g. density), dynamics (e.g. average age of its member articles) or context (e.g. the journals which the articles---group members---come from).
In total, we used over 100 features, some of which were already known to the literature, whereas the others focusing on the dynamics and context are the new, unique features proposed in this paper.
Indeed, when we rank the most valuable features contributing to successful prediction of knowledge evolution events, the new features are among the best ones.
In order to be able to perform prediction of future group changes, we have to track and learn the model on the historical cases.
For that purpose, the group changes from the past (historical evolution) need to be defined and discovered using the methods successfully applied to the Social Network Analysis field, e.g. the GED method \cite{Brodka2013a}, Tajeuna \emph{et al.} method \cite{Tajeuna2015} or other \cite{Saganowski2017a}.
Most of the methods consider the similarity between the groups in the consecutive time windows as a major factor to match similar groups and further to identify the evolution event type between them.
In our work, we apply the GED method, which facilitates both the group quantity (the number of common members) and the group quality (the importance of common members), in order to match related groups.
This allows us to enrich the co-citation evolution network with information about member relations, which is depicted in the Social Position measure \cite{Brodka2009}.

In this study, we extract groups---topical clusters (TCs)---from the bibliographic coupling networks (BCNs) and independently from the co-citation networks (CNs) for the period 1981-2010.
Next, the GED method is utilized to label four types of evolution events (changes of TCs): continuing, dissolving, merging and splitting.
Then, we use an auto-adaptive mechanism to find the most predictive machine learning model together with its parameters for each network.
Additionally, two scenarios were considered for each network: when the number of events of each kind is imbalanced (the original case) and balanced by equally sampling.
In general, the prediction quality was satisfactory good for all event types, with F-measures substantially exceeding 0.5.
Such values are significantly greater than the baseline F-measures as of 0.14--0.21 for both networks.
The feature ranking tells us that the most informative features are context-based like the number of PRE, PRB, and RMP papers belonging to the group, and the structural features like the degree, closeness, and betweenness.
While looking more carefully at the betweenness of papers from two \emph{merging} TCs, we find the significantly higher betweenness for papers that are linked across these two TCs than those connected inside the TCs.
No such enhancement in betweenness was found for \emph{continuing} TCs, while a significant decrease in average betweenness was found for \emph{splitting} TCs.
In summary, our findings suggest that evolutionary events in the landscape of physics research can be predicted accurately using various machine learning models, and understanding this predictive power in terms of important features is a worthwhile future research direction.

\section*{Results}

\subsection*{Physics research evolution for 1981-2010}

We begin with studying how scientific knowledge evolved in terms of communities of research papers, and how these communities changed over time.
There are several studies on evolution of knowledge within the set of whole journals \cite{Rosvall2010}, which is considered as the analysis on the macroscopic level.
Also some research has been carried out for the collection of papers, usually involving some subjective criterion provided by the authors, e.g. only papers cited at least 100 times\cite{Chen2010}.
As a result, they focus only on a small subset---the most prominent, frequently cited papers, which do not represent the whole diverse domain knowledge.
This kind of analysis is considered as microscopic.
In our approach, we assume that the most informative way is to analyse neither the entire journal, nor the most cited papers, but whole communities of closely related papers.
These communities emerge naturally since they share the same citation patterns.
The analysis at such level provides better balance between high and low granularity.
We call this kind of analysis as mesoscopic, because it is in-between the macroscopic scale of journals and the microscopic scale of individual papers.
However, if we perform community detection directly on the citation network, we might end up with communities consisting of both old and recent papers simultaneously.
In such case, it is difficult to interpret how scientific knowledge has evolved from the past to the present.
We should be able to explain that such and such communities represent scientific knowledge from an earlier year, whereas the other communities correspond to scientific knowledge from another consecutive year.
This enable us to compare them and to distill a picture of how scientific knowledge has evolved from past to present.
It requires, however, to construct the networks from research papers that are published in a given year (bibliographic coupling), or papers that are cited in a given year (co-citation).
The bibliographic coupling network (BCN) reflects the relation between present publications while the co-citation network (CN) represent the relation between papers which have strong influence on recent publications.
In this way, we can detect communities over the years, and study how they evolve year by year, see the Methods section for details on BCN and CN.

After building BCN and CN, the Louvain method was used to extract the community structures.
By checking the Physics and Astronomy Classification Scheme (PACS) numbers of the papers in these communities, we have shown that the BCN communities are meaningful and reflect the real structure of the scientific communities \cite{Liu2017a}.
We found indeed that papers in the same community are really focus on the closely related topics.
For the CN communities, this validation is tricky because of two problems: (i) the old Physics Review papers have no PACS numbers, and (ii) PACS was revised several times, so the same numbers in different versions can potentially refer to different topics, or the same topics are referred to by different numbers in different versions.
Nevertheless, systematic validation seems to be impossible although a quick check on some CN communities after 2010 suggests that CN community structure also reliably reflects the actual scientific community.
We refer to these validated units of knowledge evolution as topical clusters (TCs) in this paper.

\begin{figure}
  \centering
  \includegraphics[angle= 90, origin=c, width=0.6\linewidth]{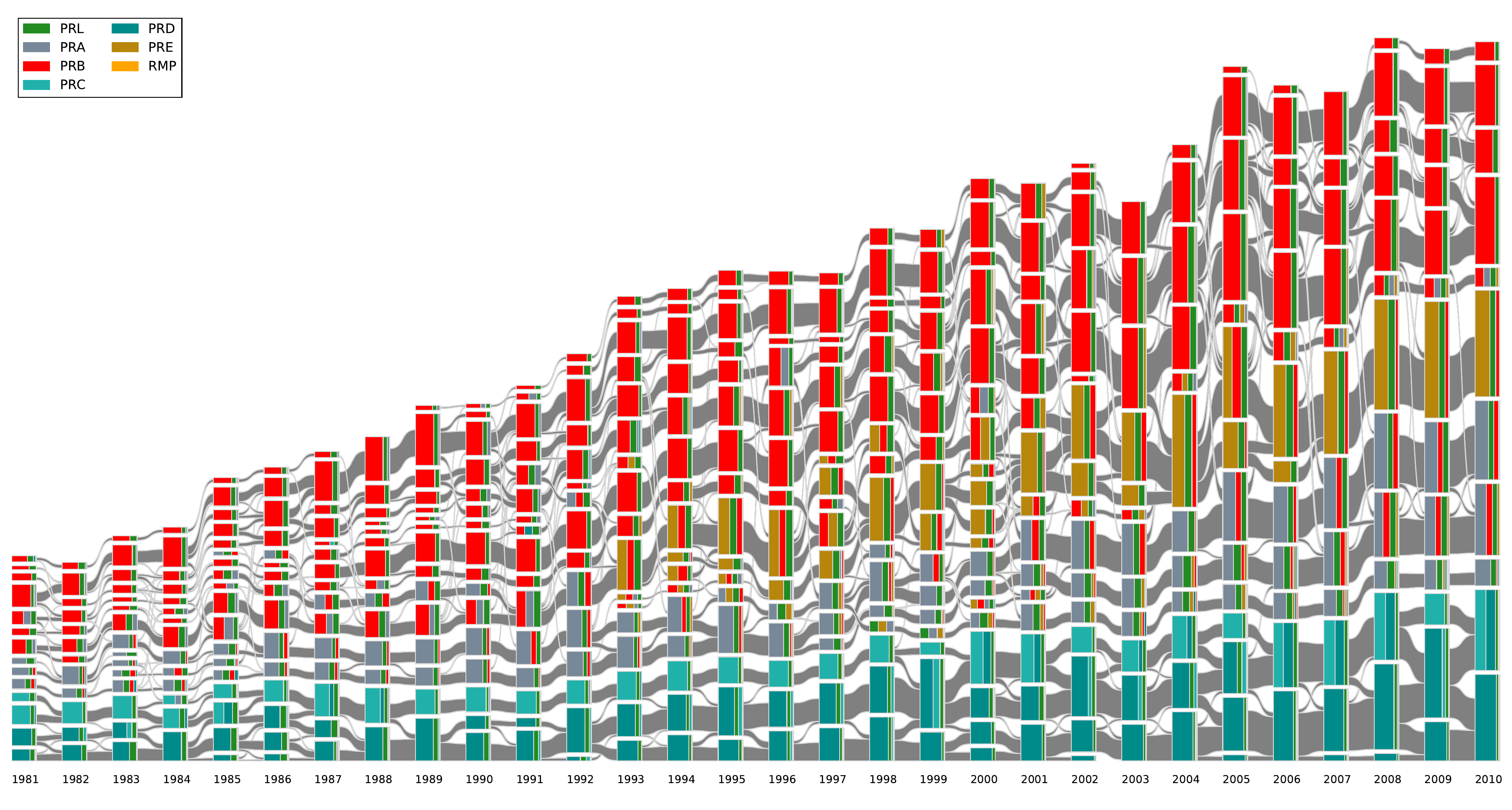}
  \caption{The alluvial diagram of APS papers from 1981 to 2010 for the BCNs. Each block in a column represents a TC and the height of the block is proportional to the number of papers in the TC. For clarity reason only TCs comprising more than 100 papers are shown. TCs in successive years are connected by streams whose widths at the left and right ends are proportional to the forward and backward intimacy indices. The colours inside a TC represent the relative contributions from different journals.}\label{fig:Fig1}
\end{figure}

In \autoref{fig:Fig1}, we provide the alluvial diagram that depicts the evolution of TCs within the BCNs for the period between 1981 to 2010.
The equivalent alluvial diagram for the CNs is shown in \autoref{fig:FigS1} in Supplementary Information (SI).
In both alluvial diagrams, we visualized the sequences of TCs, their inheritance relations, which can be intimacy indices (for the BCN communities), fraction of common members or inclusion measures (for the CN communities), and the evolution processes they undergo, see the Methods section for more details.
The events (changes) that we can discern from the alluvial diagram (shown in \autoref{fig:Fig1}) are analogous to those recognized in social group evolution\cite{Palla2007}.
They represent forming, dissolving, growing, shrinking, merging and splitting.
We found in Liu \emph{et al.} that the prediction of such events is hard, since the correlation between them is nonlinear and complex.
This challenge is addressed in the following section by tapping into the power of machine learning.

\subsection*{Event labelling}
The GED method takes into account the size and the similarity between groups (TCs) in the consecutive time frames in order to label groups change (assign event type).
There are four events considered in this work:

\begin{itemize}
\item \emph{continuing}---a research field is said to be continuing when the problems identified and solutions obtained from one year to another are of an incremental nature. It is likely correspond to the repeated hypothesis testing picture of the progress of science proposed by Karl Popper \cite{Popper1999}.
    Therefore, in the CN, this would appear as a group of papers that are repeatedly together cited year by year.
    In the BCN, this shows up as groups of articles from successive years sharing more or less the same reference list.
\item \emph{dissolving}---a research field is thought to disappear in the following year if the problems are solved or abandoned, and no new significant work is done after this.
    For the CN, we will find a group of papers that are cited up to a given year, but receives very few new citations afterwards.
    In the BCN, no new relevant papers are published in the field, hence, the reference chain terminates.
\item \emph{splitting}-–-a research field splits in the following year, when the community of scientists who used to work on the same problems, start to form two or more sub-communities, which are more and more distant from one another.
    In terms of the CN, we will find a group of papers that are almost always cited together up till a given year, breaking up into smaller and disjoint groups of papers that are cited together in the next year.
    In the BCN, we will find the transition between new papers citing a group of older papers to new papers citing only a part of this reference group.
\item \emph{merging}-–-multiple research fields are considered to have merged in the following year when the previously disjoint communities of scientists found mutual interest in each other’s field so that they solve the problems in their own domain using methods from another domain.
    In the CN, we find previously distinct groups of papers that are cited together by papers published after a given year.
    In the BCN, newly published papers will form a group commonly citing several previously disjoint groups of older papers.
\end{itemize}

The GED method has two main parameters (alpha and beta), which are the levels of inclusion that groups in the consecutive years have to cross in order to be considered as matching groups.
We have applied the GED method with the wide range of these parameters from 5\% to 100\%.
The characteristics of the considered networks required us to set the alpha and beta thresholds to very low values, i.e. 30\% for the BCN, and 10\% for the CN, see SI for more details.
In total, we have obtained 479 various events for the BCN, and 492 events for the CN, which are our observations and the labels in the prediction part of our study.
In both networks, the events distribution was imbalanced with the continuing event dominating over all other types, see \autoref{fig:Fig2}A1 and \autoref{fig:Fig2}B1.

\subsection*{Future events prediction}
The machine learning approach to prediction requires dividing the data into two parts: the training data set and test data set.
The training data is used to learn classifier, which can then label events in the test data.
The labelled values are compared with the event labels and the prediction performance is calculated.
More than 450 observations were used to train the classifiers.
Each observation contained 77 features (preselected from the initial 100) divided into three categories: microscopic features (related to nodes in the group, e.g. node degree), mesoscopic features (related to the entire group, e.g. the group size), and macroscopic features (related to the whole network, e.g. network density).
Mesoscopic features calculated for individual nodes are commonly aggregated for all nodes from the group, e.g. average node degree or betweenness in the group.
See the SI for the complete list of features used.
To automatically select the best classification algorithm (model) as well as its hyper-parameter settings to maximize the prediction performance, the Auto-WEKA software package \cite{Kotthoff2017} was utilized.
For each network, we ran the Auto-WEKA for 48 hours, which allowed us to validate nearly 20,000 configurations per network.
The metric being maximized was the F-measure, commonly used for multi-class classification.
The overall classification quality was calculated as the average F-measure for all event types, treating them as equally important.

\begin{figure}
  \centering
  \includegraphics[width=\textwidth]{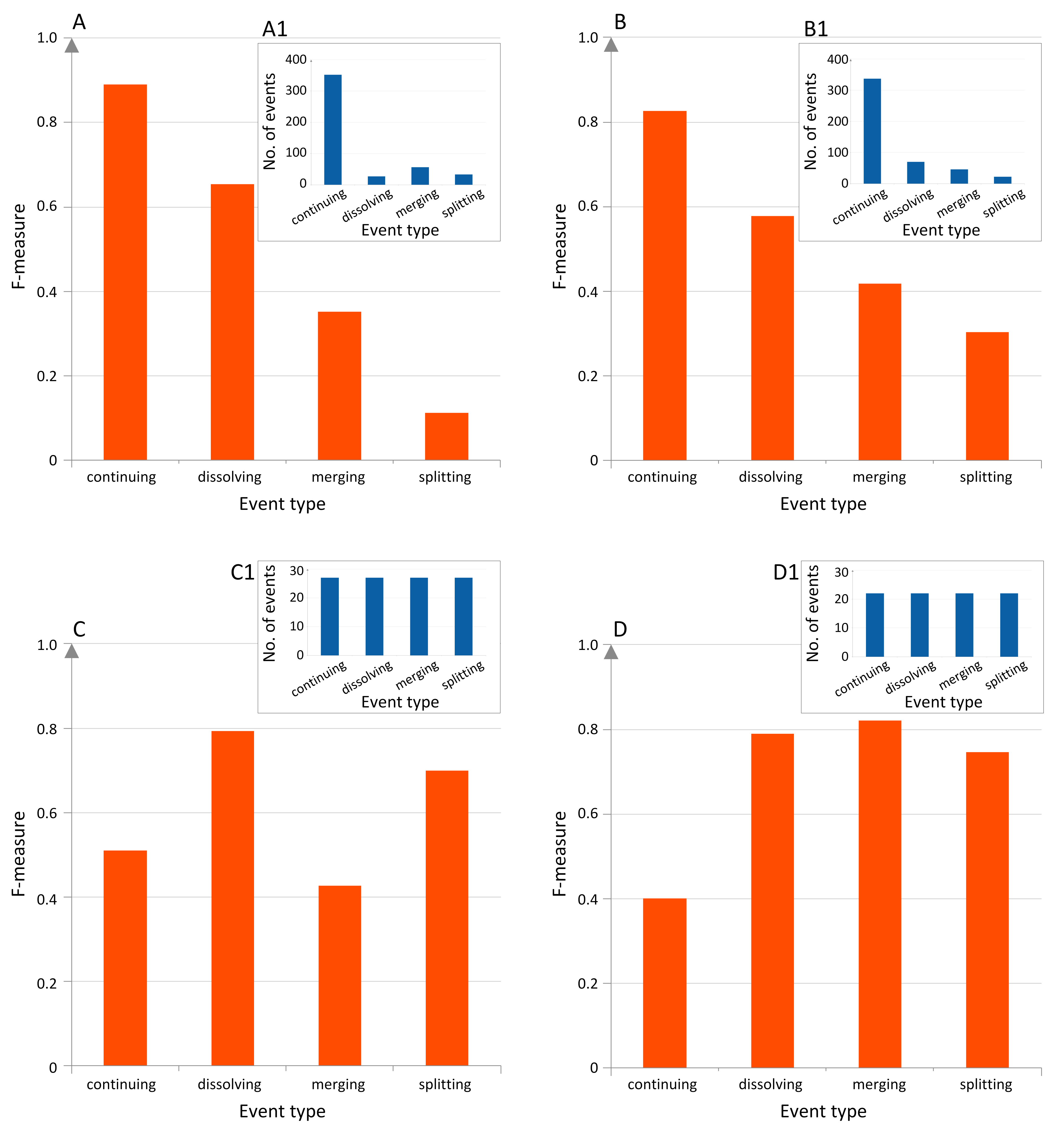}
  \caption{The prediction quality of classification results. The F-measure values for the imbalanced BCN (A) and CN (B) data sets, as well as the balanced BCN (C) and CN (D) data sets. The distribution of classes in the training sets are provided for each data set: A1, B1, C1, D1, respectively. For the imbalanced data sets, the classifier focused on the dominating continuing event. Balancing the data sets increased the overall prediction quality by over 20\%.}\label{fig:Fig2}
\end{figure}

The predicted output variable (event labels) had an imbalanced distribution.
Commonly, classifiers tend to focus on the dominant event type (class), which is very well predicted, but at the expense of the minority event types.
For the imbalanced BCN data set, the best performance was achieved with the Attribute Selected Classifier (with the SMO as base classifier), which performs feature selection\cite{Platt1999}.
The percentage of the correctly classified instances was 80.6\%, while the average F-measure was only 0.50 due to classifier focusing on continuing, which was the most frequently occurring event type, see \autoref{fig:Fig2}A.
For this event, the F-measure value was equal to 0.89, and only 7 events out of 352 were incorrectly classified.
The worst classified was the splitting event, whose F-measure was only as of 0.11.
Most of the splitting events were incorrectly classified as continuing (31 out of 33 events).
The second worst was merging, with F-measure 0.35.
Again, the majority of the merging events were wrongly classified as continuing events: 38 out of 56.
Interestingly, the splitting and merging events were never cross-classified mistakenly.
For the imbalanced CN data set, the best performance was achieved with a lazy classifier, which uses locally weighted learning \cite{Christopher1997}.
The percentage of the correctly classified instances was 73.3\%, while the average F-measure was only 0.53, again due to the classifier concentrating on the dominating continuing event type, see \autoref{fig:Fig2}B.
The F-measure value for the continuing event was only 0.83, however, as many as 50 continuing events (out of 337) were wrongly classified as dissolving.
Alike to BCN, many splitting and merging events were incorrectly classified as continuing: 17 out of 22 events, and 24 out of 46 events, with F-measure equal to 0.30 and 0.42, respectively.

By balancing the imbalanced training data sets (i.e. by equally sampling them), we force the classifiers to pay more attention to the features rather than to the number of occurrences of the particular majority event type.
As a result of balancing data sets, the previously minor event types (dissolving, merging, and splitting) were predicted much better, but with a significant drop in performance of the continuing event classification.
More importantly, by balancing the data sets we increased the overall prediction quality by over 20\%.
For the balanced BCN data set, the best performance was achieved by means of the boosting-based classifier AdaBoost with the Bayes Net as the base model.
The percentage of the correctly classified instances was 62.0\% and the average F-measure was 0.61.
The biggest sources of errors were merging events, which were wrongly classified as continuing and dissolving, as well as continuing wrongly classified as splitting.
The best classified event was dissolving (only 4 mistakes in 27 classifications, the overall score 0.79) followed by the splitting event (6 mistakes in 27 classifications, overall F-measure 0.70).
For the balanced CN data set, the Attribute Selected Classifier (with the PART as base classifier) provided the best results---the percentage of the correctly classified instances was 69.32\%, while the average F-measure was 0.69\cite{Frank1998}.
The dissolving, merging, and splitting events were classified very well with the F-measure values equal to 0.79, 0.82, and 0.75 respectively. Most of the continuing events were wrongly classified as splitting (13 out of 22), which resulted in lower F-measure value 0.40.

What is interesting for us to note is that the prediction results for the CN being slightly better than for the BCN.
A possible explanation is that for the CN we used a richer similarity measure containing users importance information.
Thus the event tracking and therefore the ground truth could be more accurate.
Overall, the prediction quality expressed by the average F-measure was very good for the imbalanced as well as for the balanced data sets, as the baseline results obtained with the ZeroR classifier were much worse: F-measure 0.21 for both, BCN and CN, imbalanced data sets, 0.18 for the balanced BCN and 0.14 for the balanced CN.
For each data set different classifier turned out to be the best, however most models were wrapped with the boosting or meta classifiers.

\subsection*{Predictive feature ranking}

\begin{figure}
  \centering
  \includegraphics[width=\textwidth]{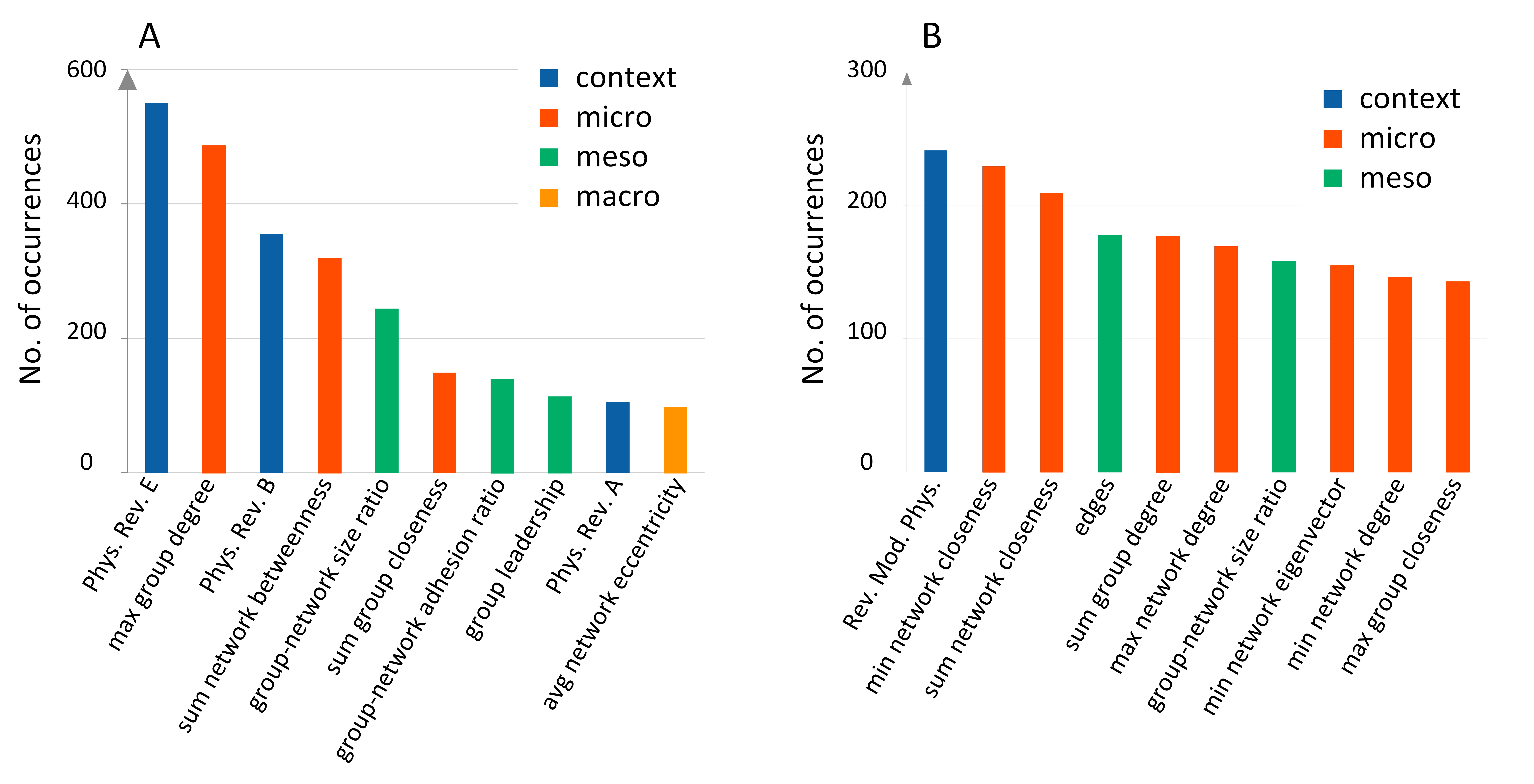}
  \caption{Feature ranking. The most frequently selected features in 1000 iterations for the BCN (A) and CN (B) data sets. The context-based features (number of papers published in a given journal) turned out to be the most informative, followed by the microscopic structural measures, especially closeness, degree and betweenness.}\label{fig:Fig3}
\end{figure}

The feature selection technique is used in machine learning to find the most informative features, to avoid classifier overfitting, to eliminate (or at least to reduce) the noise in the data as well as to provide some explanations about phenomena\cite{Yang1998}.
By repeating the feature selection 1000 times, we obtained 1000 sets of selected features.
Next, we calculated how many times each feature has been selected, thus, receiving the ranking of the most often selected features.
For the BCN, the context-based features dominated the ranking.
It referred, especially the number of papers from the Physical Review E, Physical Review B and Physical Review A, see \autoref{fig:Fig3}A.
Beside the context, the network features based on degree, betweenness, size and closeness measures were most informative, which tells us that the structural properties are as important as context awareness.
The context-based feature, i.e. the number of papers published in Review of Modern Physics, was the most often selected for the CN data set.
It is followed by closeness- and degree-based features in the ranking, see \autoref{fig:Fig3}B.
For both networks macroscopic features were ranked rather low, which suggests that the general network profile is not very important, perhaps because of the smooth changes in the entire network.
Surprisingly, the dynamic features, e.g. related to the average age of references (for BCN) and age of articles (for CN) did not show informative value and were ranked very low for both networks.
The rankings were validated in the additional two years of data available (2010-2012).
The prediction was performed twice: (i) using all features, and (ii) using the top 10 ranked features only.
Selecting only the top 10 features, boosted the quality of the prediction by 11\% for the CN, and by 2\% for the BCN, which underlines the necessity of the feature selection process.

\subsection*{Changes to the Betweenness Distributions Associated with Merging and Splitting Events in BCN}

\begin{figure*}[t!]
  \centering
  \begin{subfigure}[t]{0.24\textwidth}
    \centering
    \includegraphics[width=\textwidth]{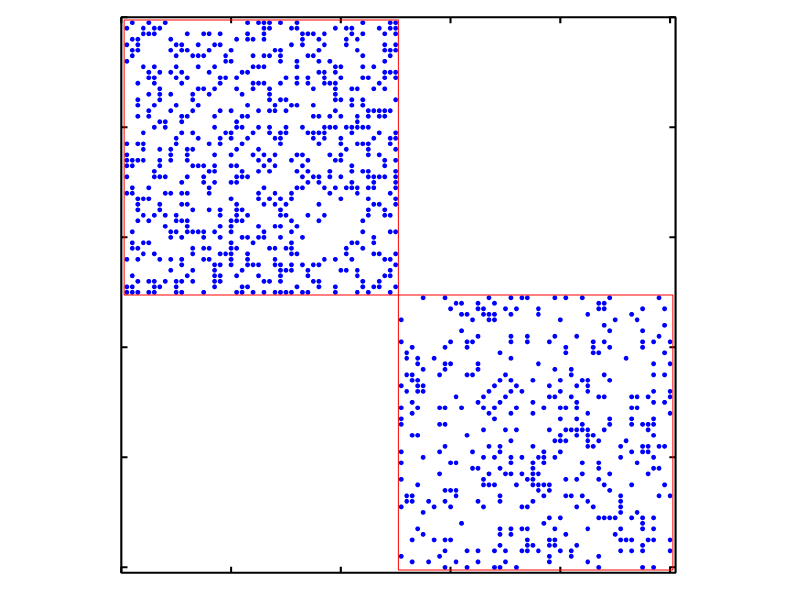}
    \caption{}
  \end{subfigure}%
  ~
  \begin{subfigure}[t]{0.24\textwidth}
    \centering
    \includegraphics[width=\textwidth]{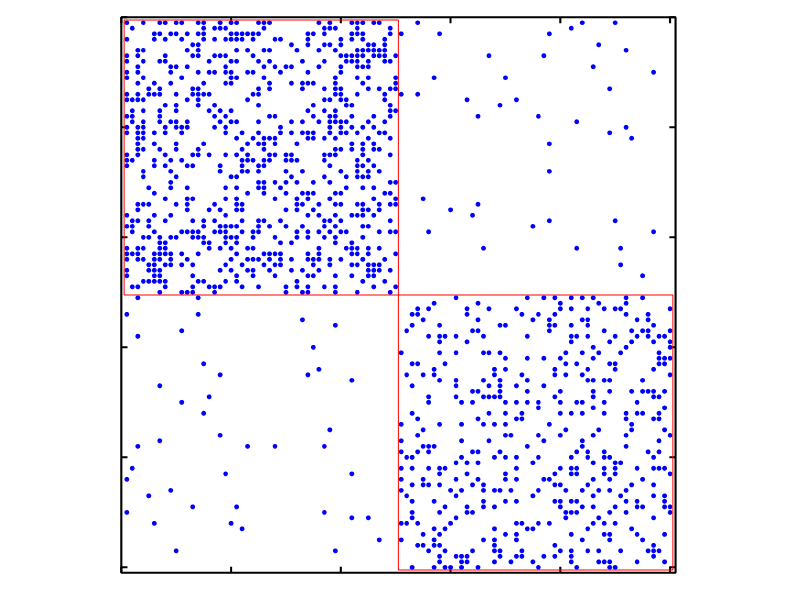}
    \caption{}
  \end{subfigure}%
  ~
  \begin{subfigure}[t]{0.24\textwidth}
    \centering
    \includegraphics[width=\textwidth]{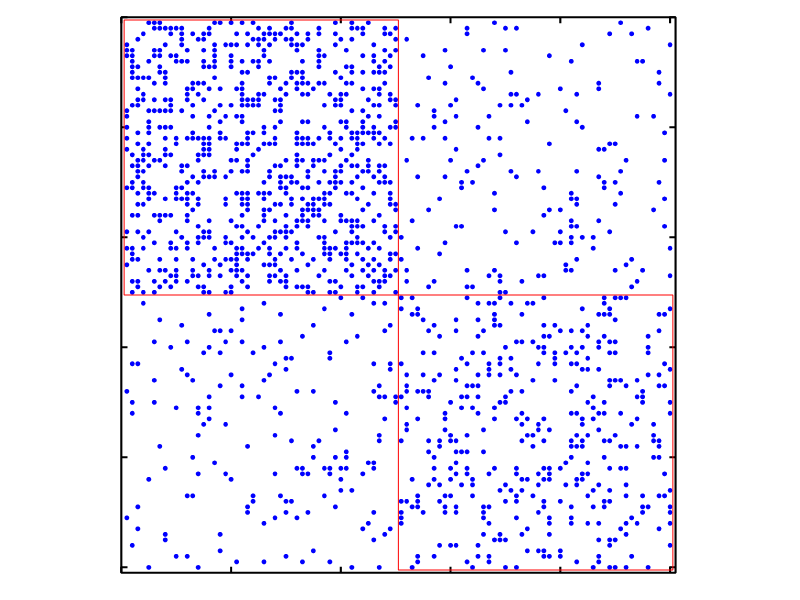}
    \caption{}
  \end{subfigure}%
  ~
  \begin{subfigure}[t]{0.24\textwidth}
    \centering
    \includegraphics[width=\textwidth]{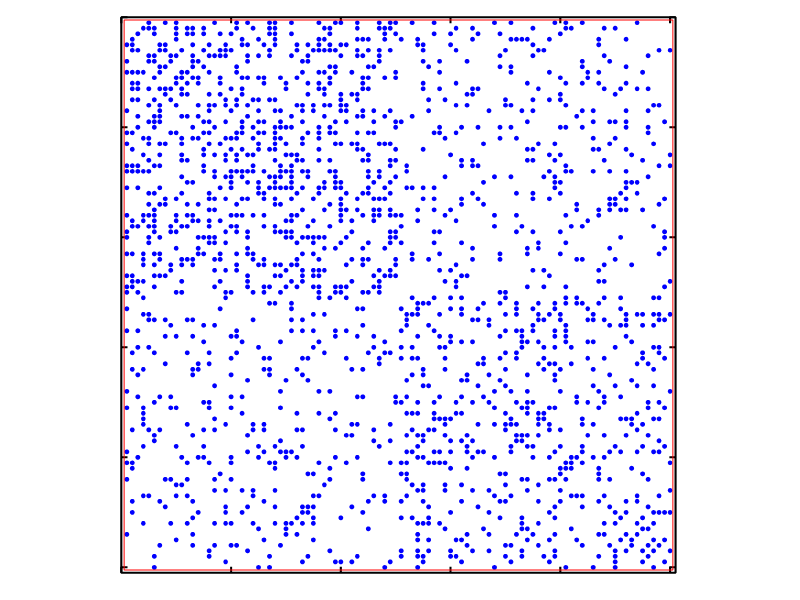}
    \caption{}
  \end{subfigure}
  \caption{Part of the BCN adjacency matrix for two TCs (red boxes) that ultimately merged. (A) No links between the two TCs at first. (B) Few links between the two TCs. (C) More links between the two TCs. (D) Many links between the two TCs, leading to their identification as a single merged TC (big red box) by the Louvain method.}\label{fig:Fig4}
\end{figure*}

Having the list of best predictive features, \autoref{fig:Fig3}, we can analyse some of them more in-depth to look for early warning signals.
Basically, we believe that scientific knowledge evolves slowly, and this slow evolution drives the evolution of citation patterns.
Therefore, there must be specific changes in citation patterns that precede merging and splitting events.
Besides the number of PRE papers in a TC, the sum\_network\_betweenness is also a strongly predictive feature, see \autoref{fig:Fig3}A.
This suggests that we should look at the betweenness of papers in the BCN more carefully.
The betweenness of the node denotes what percentage of shortest paths between all pairs of nodes in the network passes a given node.
Values of nodes' betweenness can be aggregated (sum, average, max, min) for all nodes in the TC, as what we list in \autoref{tab:TabS1}.
However, in this section we only focus on the distribution of original node betweenness.
Naively, when we consider the part of the BCN adjacency matrix corresponding to two TCs that ultimately merged, we expect to find few links between TCs at first.
But as the number of links between TCs increase over time, the modularity-maximizing Louvain method will eventually merge the two TCs into a single TC.
This is shown schematically in \autoref{fig:Fig4}, where in general betweenness will increase on average with time as the two TCs merge.
In reality, there are always links between TCs, and the numbers and strengths of these links fluctuate over time.
To develop a more quantitative description of the merging events outlined in \autoref{fig:Fig1}, as well as splitting and continuing events, we focus on five events going from 1999 to 2000, shown in \autoref{tab:events}.

\begin{table}
\centering
\begin{minipage}[t]{0.5\textwidth}
\vspace{0pt}
\includegraphics[width=\textwidth]{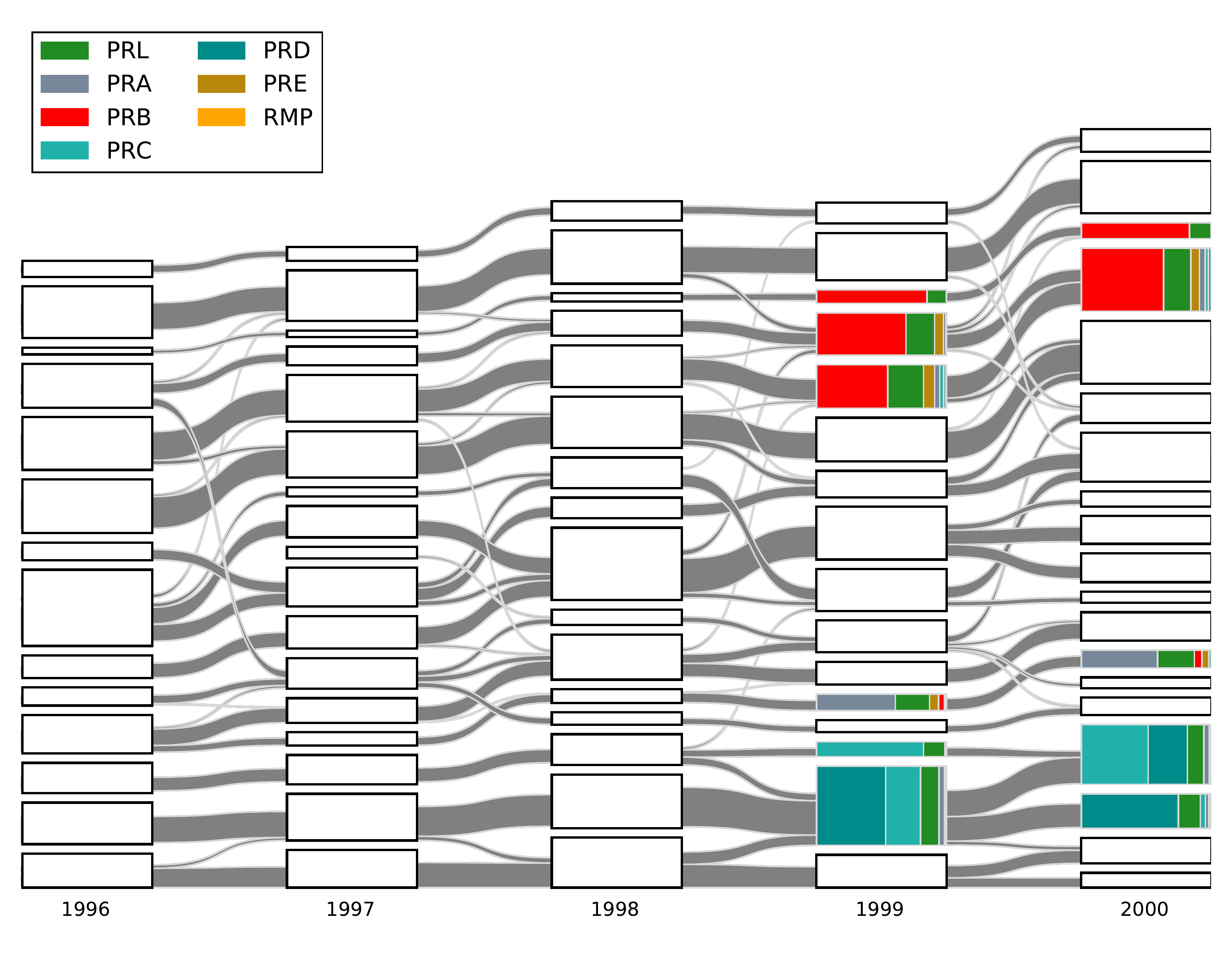}
\end{minipage}%
\begin{minipage}[t]{0.5\textwidth}
\vspace{13.5ex}
\centering
  \begin{tabular}{|c|c|c|}
    \hline
    \textbf{TC in 1999} & \textbf{event} & \textbf{TC in 2000}\\
    \hline
    1999.01	& split	& 2000.02, 2000.03 \\
    \hline
    1999.01, 1999.02 &	merge &	2000.03 \\
    \hline
    1999.04	& continue &	2000.06 \\
    \hline
    1999.11, 1999.12 &	merge &	2000.15 \\
    \hline
    1999.13	& continue &	2000.16 \\
    \hline
  \end{tabular}
\end{minipage}
\caption{The five evolution events from 1999 to 2000 in the BCN alluvial diagram \autoref{fig:Fig1} that we will study quantitatively. The naming convention for TC is that four digits before `.' is the year of TC, two digits after `.' is the position of the TC in the diagram, starting with 00 for the bottom TC, the one just above bottom is 01 and so on. In the left panel, we highlight the related TCs.}\label{tab:events}
\end{table}

\subsubsection*{$\mathbf{1999.01} + \mathbf{1999.02} \rightarrow \mathbf{2000.03}$}
\begin{figure*}[t!]
  \centering
  \begin{subfigure}[t]{0.45\textwidth}
    \centering
    \includegraphics[width=\textwidth]{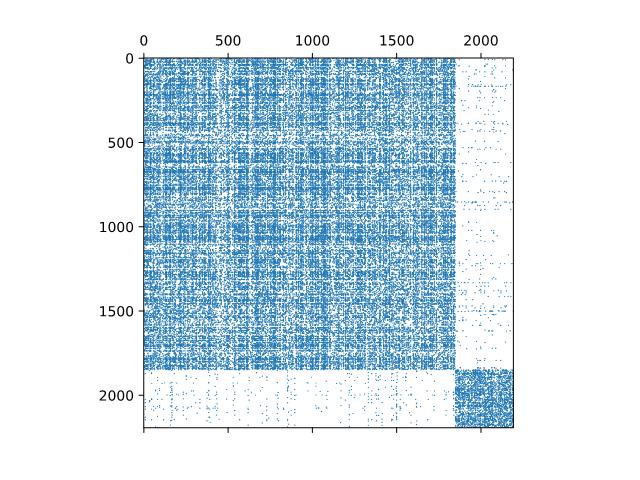}
    \caption{}
  \end{subfigure}%
  ~
  \begin{subfigure}[t]{0.45\textwidth}
    \centering
    \includegraphics[width=\textwidth]{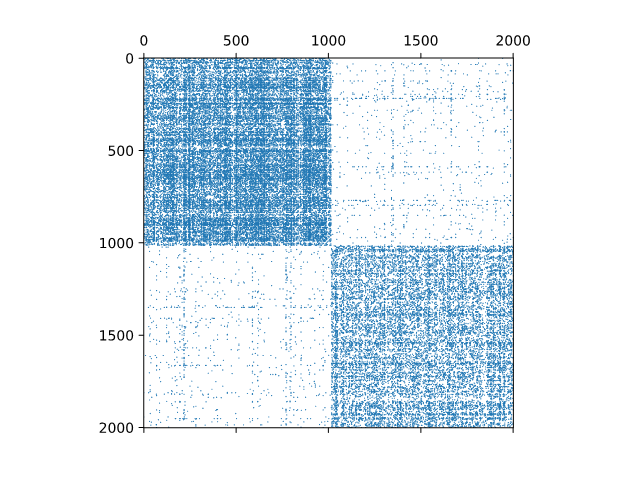}
    \caption{}
  \end{subfigure}
  \caption{(A)The adjacency matrix of the BCN associated with the TCs 1999.01 (top dense block) and 1999.02 (bottom dense block). (B)The adjacency matrix of the BCN associated with the TCs 1999.11 (top dense block) and 1999.12 (bottom dense block).}\label{fig:Fig5}
\end{figure*}

Let us consider the part of the BCN associated with the TCs.
For example, for 1999.01 and 1999.02, we can see from \autoref{fig:Fig5}(A) that connections within 1999.01 and 1999.02 are very dense, but there are also some links between the two TCs.
In fact, we find 164 out of 1849 papers in 1999.01 with non-zero bibliographic coupling to 144 papers in 1999.02 (344 papers).
The natural question we then ask is: are the betweenness of the 164 papers in 1999.01 that are coupled to 1999.02 larger, equal, or smaller than the betweeness of the rest 1685 papers in 1999.01 not coupled to 1999.02?
Alternatively, if we think of the 164 papers as randomly sampled from the 1849 papers in 1999.01, are we sampling the 164 betweenness in an unbiased fashion?
To distinguish the different parts of the TC, we call all papers in 1999.01 which have coupling with papers in 1999.02 as $1999.01a$, and the rest of papers as $1999.01b$.
For more detail analysis, we will divide $1999.01a$ and $1999.01b$ into $1999.01a\alpha$, $1999.01a\beta$, $1999.01b\alpha$, $1999.01b\beta$.
$1999.01a\alpha$ consist of 17 papers in $1999.01a$ that do not have references in common with papers in $1999.01b$, $1999.01a\beta$ consist of 147 papers in $1999.01a$ that have references in common with papers in $1999.01b$, $1999.01b\alpha$ are 907 papers in $1999.01b$ that have references in common with papers in $1999.01a$ and $1999.01b\beta$ represents 778 papers in $1999.01b$ that do not have references in common with papers in $1999.01a$.

\begin{table}
  \centering
  \begin{tabular}{|c|c|c|c|}
    \hline
     \multirow{2}{*}{} & \multicolumn{3}{c|}{percentile} \\
    \cline{2-4}
      & 25 & 50 & 75 \\
    \hline
    1999.01 & \num{8.06d-6} & \num{5.73d-5} & \num{2.05d-4} \\
    \hline
    $1999.01a$ & \num{5.90d-5} & \num{1.58d-4} & \num{4.67d-4} \\
    \hline
    $1999.01a\alpha$ & \num{7.77d-6} & \num{1.95d-5} & \num{2.44d-4} \\
    \hline
    $1999.01a\beta$ & \num{5.29d-6} & \num{4.96d-5} & \num{2.48d-4} \\
    \hline
    $1999.01b$ & \num{6.22d-6} & \num{5.04d-5} & \num{1.88d-4} \\
    \hline
    $1999.01b\alpha$ & \num{8.59d-6} & \num{6.00d-5} & \num{2.14d-4} \\
    \hline
    $1999.01b\beta$ & \num{7.97d-6} & \num{5.32d-5} & \num{1.83d-4} \\
    \hline
    $1999.02$ & \num{2.47d-6} & \num{5.54d-5} & \num{2.13d-4} \\
    \hline
    $1999.02a$ & \num{3.08d-5} & \num{1.13d-4} & \num{3.17d-4} \\
    \hline
    $1999.02b$ & \num{2.14d-7} & \num{1.44d-5} & \num{1.60d-4} \\
    \hline
    1999.11 & \num{1.73d-5} & \num{9.04d-5} & \num{2.81d-4} \\
    \hline
    $1999.11a$ & \num{6.38d-5} & \num{1.98d-4} & \num{4.61d-4} \\
    \hline
    $1999.11b$ & \num{9.91d-6} & \num{6.17d-5} & \num{2.17d-4} \\
    \hline
    1999.12 & \num{6.56d-6} & \num{4.54d-5} & \num{1.62d-4} \\
    \hline
    $1999.12a$ & \num{2.74d-5} & \num{9.08d-5} & \num{2.33d-4} \\
    \hline
    $1999.12b$ & \num{2.52d-6} & \num{2.69d-5} & \num{1.20d-4} \\
    \hline
  \end{tabular}
  \caption{The 25th, 50th and 75th percentiles of the betweenness of 1849 papers in 1999.01, the 164 papers in $1999.01a$, the 17 papers in $1999.01a\alpha$, the 147 papers in $1999.01a\beta$, the 1685 papers in $1999.01b$, the 907 papers in $1999.01b\alpha$, the 778 papers in $1999.01b\beta$; the 344 papers in 1999.02, the 144 papers in $1999.02a$, and the 200 papers in $1999.02b$; the 1014 papers in 1999.11, the 299 papers in $1999.11a$, the 715 papers in $1999.11b$ and the 988 papers in 1999.12, the 347 papers in $1999.12a$, the 641 papers in $1999.12b$.}\label{tab:percentile}
\end{table}

\begin{figure*}[t!]
  \centering
  \begin{subfigure}[t]{0.3\textwidth}
    \centering
    \includegraphics[width=\textwidth]{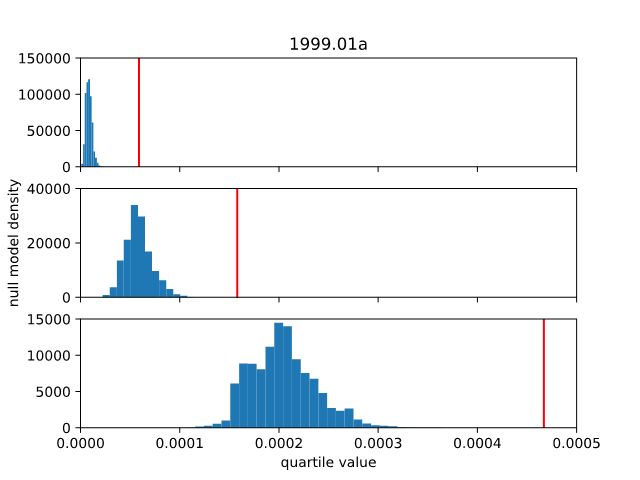}
    \caption{}
  \end{subfigure}%
  ~
  \begin{subfigure}[t]{0.3\textwidth}
    \centering
    \includegraphics[width=\textwidth]{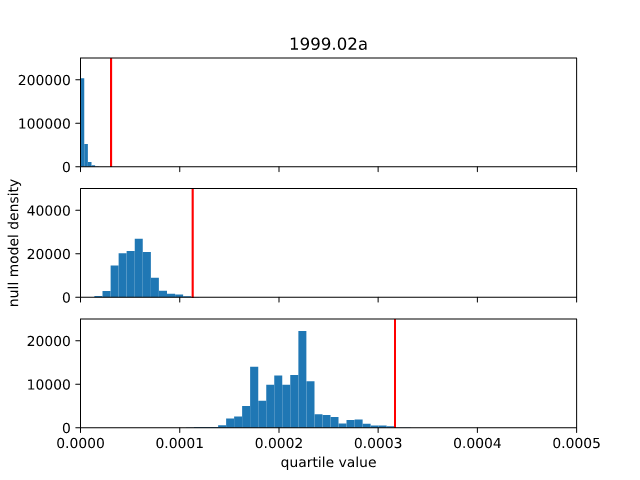}
    \caption{}
  \end{subfigure}%
  ~
  \begin{subfigure}[t]{0.3\textwidth}
    \centering
    \includegraphics[width=\textwidth]{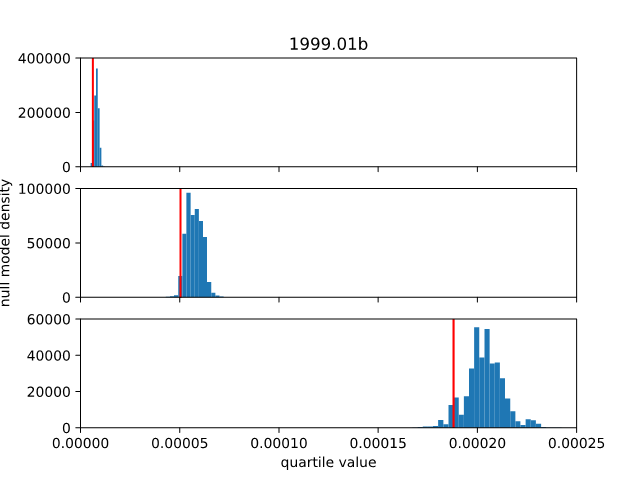}
    \caption{}
  \end{subfigure}\\
  \begin{subfigure}[t]{0.3\textwidth}
    \centering
    \includegraphics[width=\textwidth]{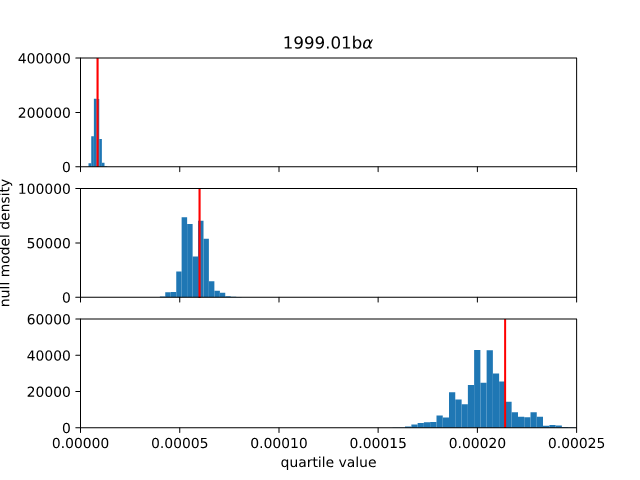}
    \caption{}
  \end{subfigure}%
  ~
  \begin{subfigure}[t]{0.3\textwidth}
    \centering
    \includegraphics[width=\textwidth]{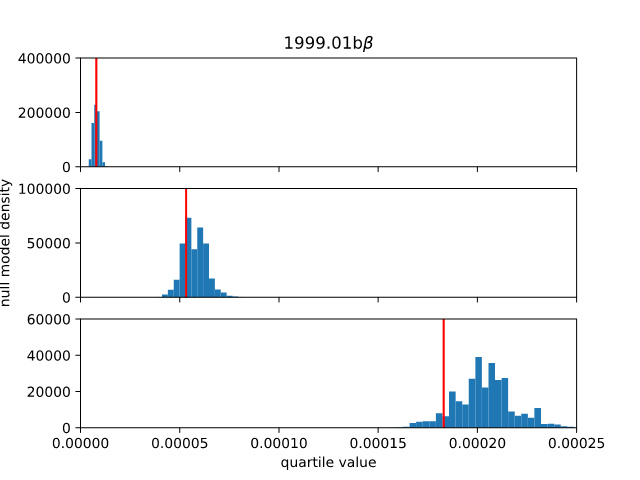}
    \caption{}
  \end{subfigure}%
  ~
  \begin{subfigure}[t]{0.3\textwidth}
    \centering
    \includegraphics[width=\textwidth]{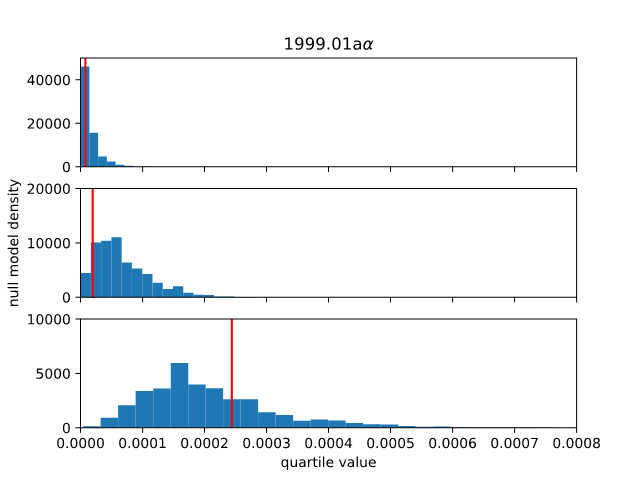}
    \caption{}
  \end{subfigure}\\
  \begin{subfigure}[t]{0.3\textwidth}
    \centering
    \includegraphics[width=\textwidth]{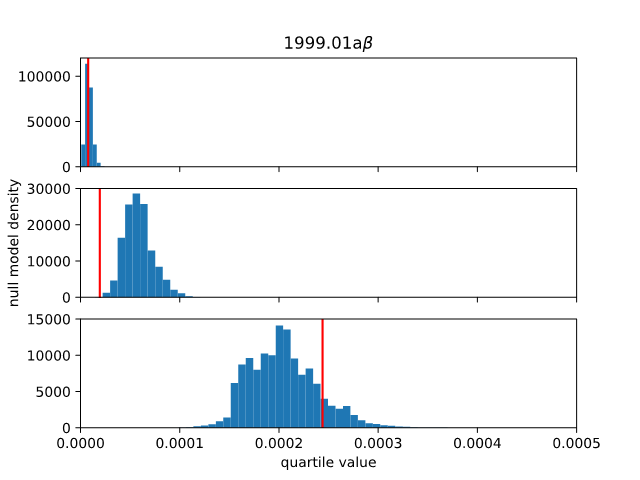}
    \caption{}
  \end{subfigure}%
  ~
  \begin{subfigure}[t]{0.3\textwidth}
    \centering
    \includegraphics[width=\textwidth]{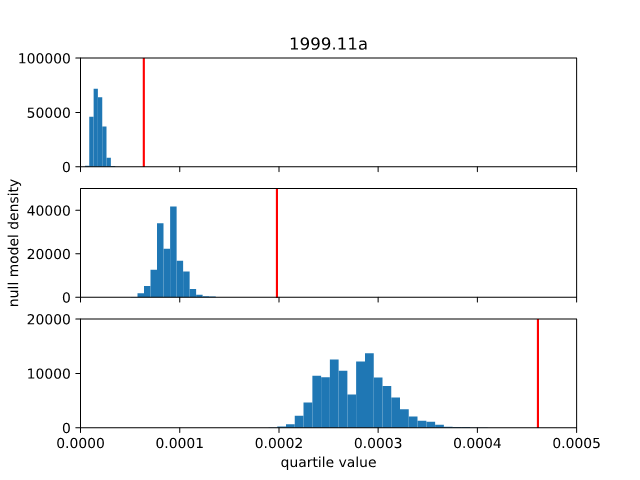}
    \caption{}
  \end{subfigure}%
    ~
  \begin{subfigure}[t]{0.3\textwidth}
    \centering
    \includegraphics[width=\textwidth]{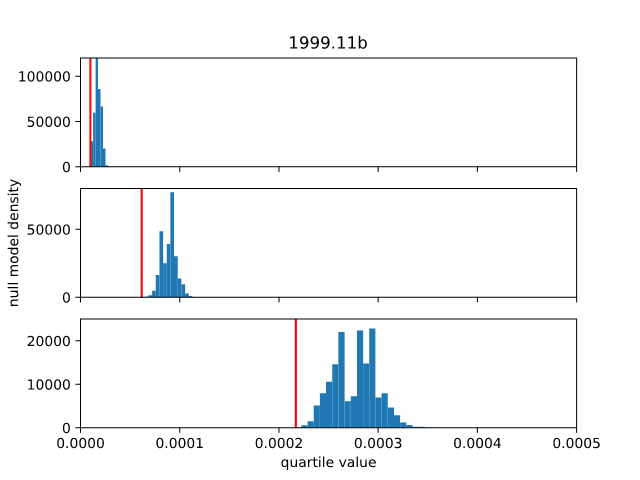}
    \caption{}
  \end{subfigure}
  \\
  \begin{subfigure}[t]{0.3\textwidth}
    \centering
    \includegraphics[width=\textwidth]{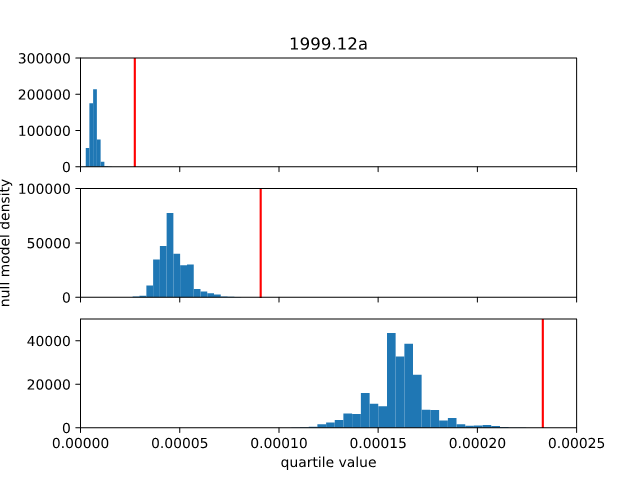}
    \caption{}
  \end{subfigure}%
    ~
  \begin{subfigure}[t]{0.3\textwidth}
    \centering
    \includegraphics[width=\textwidth]{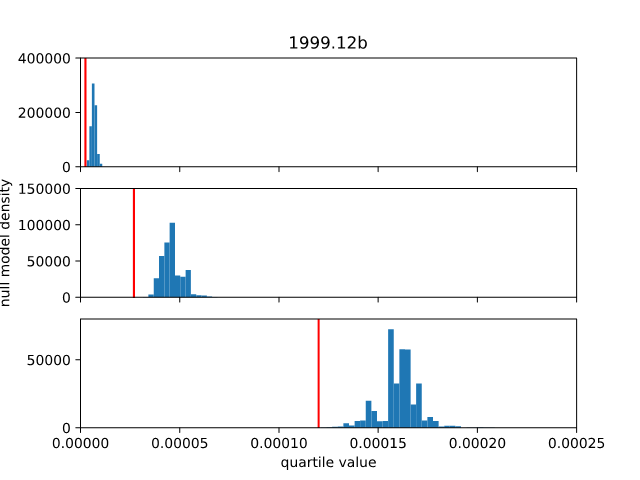}
    \caption{}
  \end{subfigure}%
  \caption{The lower (top), median (middle), and top quartile (bottom) of the betweennesses in (A) $1999.01a$, (B) $1999.02a$, (C) $1999.01b$, (D) $1999.01b\alpha$, (E) $1999.01b\beta$, (F) $1999.01a\alpha$, (G) $1999.01a\beta$, (H) $1999.11a$, (I) $1999.11b$, (J) $1999.12a$, (K) $1999.12b$ shown as red vertical lines, and \num{e6} random samples of the same number of betweennesses from 1999.01 (A, C, D, E, F, G) or 1999.02 (B) or 1999.11 (H, I) or 1999.12 (J, K) shown as blue histograms.}\label{fig:Fig6}
\end{figure*}

In \autoref{tab:percentile}, we show the 25th, 50th and 75th percentiles of the papers in these smaller groups, compared to those of 1849 papers in 1999.01 and 344 papers in 1999.02.
As we can see, the 25th, 50th, 75th percentile betweenness in the connecting parts ($1999.01a$ and $1999.02a$) are all higher than the 25th, 50th, 75th percentile betweenness in the non-connecting parts ($1999.01b$ and $1999.02b$).
More importantly, these percentile betweenness are higher than the 25th, 50th, 75th percentile betweenness of the TCs 1999.01 and 1999.02 themselves.
To test how significant these quartiles are in $1999.01a$, we randomly sampled 164 betweenness values from 1999.01 \num{e6} times, and measured the quartiles of these samples.
When we draw random samples from a TC, the 25th percentile, the 50th percentile, and the 75th percentile, depends on the size of the TC.
There is more variability in these quartiles in smaller samples than they are in larger samples.
Therefore, in the test for statistical significance, the observed quartile has to be tested against different null model quartiles for samples of different sizes.
To do this, we draw samples with a range of sizes from the same set of betweenness, and for a given quartile (25\%, 50\%, or 75\%), fit the minimum quartile value against sample size to a cubic spline, and the maximum quartile value against sample size to a different cubic spline.
With these two cubic splines, we can then check whether the observed quartile value for a sample of size $n$ is more than or less than the null model minimum or maximum using cubic spline interpolation.
From the histograms shown in \autoref{fig:Fig6}(A), we see that the betweenness quartiles of $1999.01a$ are statistically larger than random samples of the same size from 1999.01, at the level of $p < \num{d-6}$, which means the papers in $1999.01a$ have significantly larger betweenness than other papers in 1999.01.

We also checked the statistical significance of the larger betweenness values of $1999.02a$, against \num{e6} random samples of the same length (144) from 1999.02.
From \autoref{fig:Fig6}(B), we can derive that the quartiles of $1999.02a$ are only a little larger than the tails of the quartile histograms of the random samples, but their statistical significance is still at the level of $p < \num{d-6}$.

\subsubsection*{$\mathbf{1999.01} \rightarrow \mathbf{2000.02} + \mathbf{2000.03}$}

When a TC splits into two in the next year, we expect the links between two parts $a$ and $b$ in the TC to have thinned out to the point that the modularity $Q$ of the whole is lower than the modularities $Q_a$ and $Q_b$ of the two parts.
However, in general, we would not know how to separate the TC into the two parts $a$ and $b$.
Fortunately, for the $1999.01 \rightarrow 2000.02 + 2000.03$ splitting event, we also know the part $1999.01a$, which merged with $1999.02a$,  became 2000.03.
Therefore, we might naively expect $1999.01b$ to be the part that split from $1999.01$ to become 2000.02.
If we test the quartiles of $1999.01b$, against random samples of the same size from 1999.01, we find the histograms shown in \autoref{fig:Fig6}(C).
As we can see, the betweenness quartiles of $1999.01b$ are quite a bit lower than the typical values in 1999.01, but this difference is statistically not as significant as the quartiles of $1999.01a$.
Thinking about this problem more deeply, we realized that while papers in $1999.01b$ have no references in common with 1999.02, some of them do share common references with $1999.01a$.
Let us call these sets of papers $1999.01a\alpha$ (papers do not have references in common with papers in $1999.01b$), $1999.01a\beta$ (papers have references in common with papers in $1999.01b$), $1999.01b\alpha$(papers have references in common with papers in $1999.01a$), and $1999.01b\beta$ (papers that do not have references in common with papers in $1999.01a$).
In \autoref{fig:Fig6}(D), we learn from the histograms that the betweenness quartiles of $1999.01b\alpha$ are indistinguishable with random samples of the same size from 1999.01.
On the other hand, from the histograms in \autoref{fig:Fig6}(E), we find out that while the lower betweenness quartile of $1999.01b\beta$ is indistinguishable with the random samples of the same size from 1999.01, its median and upper quartile are both on the low sides of the random sample distributions.
This suggests a split of 1999.01 to (1999.01a + $1999.01b\alpha$) + $1999.01b\beta$.

Just to be safe, we also checked the betweenness quartiles of $1999.01a\alpha$ and $1999.01a\beta$, against random samples of the same sizes from 1999.01.
As we can see from \autoref{fig:Fig6}(F) and (G), the lower quartiles and medians are lower than those obtained from random samples, but the upper quartiles are decidedly higher.
However, the difference between $1999.01a\alpha$ and $1999.01a\beta$ is not as obvious as difference between $1999.01b\alpha$ and $1999.01b\beta$, one possible reason is the smaller sample size (17, 147 vs. 907, 778).
Again, these results are consistent with the picture that the rise in betweenness in $1999.01a$ is driving the merging with $1999.02a$, while the fall in betweenness in $1999.01b\beta$ is driving a splitting inside 1999.01.

\subsubsection*{$\mathbf{1999.11} + \mathbf{1999.12} \rightarrow \mathbf{2000.15}$}

Although a small part split off from each of 1999.11 and 1999.12, the main event associated with the two TCs was a symmetric merging.
Looking again into the relevant parts of the BCN, we found 299 out of 1014 papers in 1999.11 coupled to 347 out of 988 papers in 1999.12, and we call them $1999.11a$ and $1999.12a$, respectively.
As we can see from the histograms in \autoref{fig:Fig6}(H) and (J), the betweenness quartiles in $1999.11a$ and $1999.12a$ are significantly higher than one would expect from random samples of 1999.11 and 1999.12.
Simultaneously, the betweenness quartiles in $1999.11b$ and $1999.12b$ are significantly lower than in random samples of 1999.11 and 1999.12 (see \autoref{fig:Fig6}(I) and (K)).
Therefore, what we are seeing here might be the early warning signals of merging, as well as that of the asymmetric splitting.

\subsubsection*{$\mathbf{1999.04} \rightarrow \mathbf{2000.06}$ and $\mathbf{1999.13} \rightarrow \mathbf{2000.16}$}

So far we have learnt that a decrease in betweenness within a TC signals a possible split, whereas an increase in betweenness of the part of the TC coupled to another TC signals a merger between the two TCs.
For this story to be consistent, we must not see these signals in the continuing events $1999.04 \rightarrow 2000.06$ and $1999.13 \rightarrow 2000.16$.
However, if we go through the full BCN, we find that 370 out of 389 papers in 1999.04 and 308 out of 319 papers in 1999.13 are coupled to papers outside of these TCs, which suggests the possibility of merging or splitting.
%
\begin{table}
  \centering
  \begin{tabular}{|c|c|c|c|c|c|c|c|c|}
    \hline
	\multirow{3}{*}{} & \multicolumn{4}{c|}{1999.04}	& \multicolumn{4}{c|}{1999.13} \\
    \cline{2-9}
     & \multirow {2}{*}{size} & \multicolumn{3}{c|}{percentile} & \multirow {2}{*}{size} & \multicolumn{3}{c|}{percentile} \\
    \cline{3-5}\cline{7-9}
     &      & 25 & 50 & 75 & & 25 & 50 & 75 \\
    \hline
    1999.00	& 12 & \num{9.0d-5} & \num{1.1d-3} & \num{2.3d-3} &	1 &	-            &	-            & \num{1.8d-3} \\
    \hline
    1999.01	& 56 & {\color{red}\num{1.6d-4}}	& {\color{red}\num{4.2d-4}} & \num{1.0d-3} &	6 &	\num{2.0d-4} & \num{4.9d-4}	 & \num{6.5d-4} \\
    \hline
    1999.02	& 6	 & \num{3.0d-4}	& \num{5.1d-4} & \num{7.4d-4} &	2 &	\num{6.0d-4} &	-	         & \num{2.6d-4} \\
    \hline
    1999.03	& 25 & \num{1.6d-5}	& \num{4.3d-4} & \num{8.1d-4} &	0 &	-            &	-            &	- \\
    \hline
    1999.04	& -	 & -            &	-          &	-         &	8 &	\num{1.5d-4} &	\num{4.8d-4} & \num{8.0d-4} \\
    \hline
    1999.05	& 179& \num{4.9d-5}	& \num{1.7d-4} & \num{4.5d-4} &	4  & \num{2.2d-4} &	\num{4.3d-4} &	\num{6.5d-4} \\
    \hline
    1999.06	& 110& \num{8.7d-5}	& \num{2.0d-4} & \num{6.2d-4} &	40 & \num{5.9d-5} &	\num{1.6d-4} &	\num{4.5d-4} \\
    \hline
    1999.07	& 29 & \num{1.7d-4}	& \num{5.6d-4} & \num{1.2d-3} &	44 & \num{1.4d-4} &	\num{3.1d-4} &	\num{5.5d-4} \\
    \hline
    1999.08	& 63 & \num{1.1d-4}	& \num{3.2d-4} & \num{8.6d-4} &	17 & \num{2.2d-4} &	\num{5.2d-4} &	\num{8.5d-4} \\
    \hline
    1999.09	& 49 & \num{7.8d-5}	& \num{2.6d-4} & \num{8.0d-4} &	99 & {\color{red}\num{8.0d-5}} &	{\color{red}\num{2.5d-4}} &	\num{4.8d-4} \\
    \hline
    1999.10	& 53 & \num{1.2d-4}	& \num{3.8d-4} & \num{8.2d-4} &	254& \num{3.6d-5} &	\num{8.8d-5} &	\num{2.7d-4} \\
    \hline
    1999.11	& 89 & \num{1.0d-4}	& {\color{red}\num{3.2d-4}} & \num{9.2d-4} &	71 & {\color{red}\num{1.4d-4}} &	{\color{red}\num{3.4d-4}} &	\num{5.7d-4} \\
    \hline
    1999.12	& 53 & \num{8.7d-5}	& \num{2.9d-4} & \num{9.3d-4} &	39 & \num{1.3d-4} &	\num{2.7d-4} &	\num{4.6d-4} \\
    \hline
    1999.13	& 9	 & \num{1.3d-4}	& \num{4.2d-4} & \num{1.1d-3} & -  & -            &	-            &	- \\
    \hline
    1999.14	& 62 & {\color{red}\num{1.4d-4}}	& {\color{red}\num{4.8d-4}} & \num{1.0d-3} &	210& \num{4.2d-5} &	\num{1.0d-4} &	\num{2.7d-4} \\
    \hline
    1999.15	& 17 & \num{1.8d-4}	& \num{3.6d-4} & \num{9.7d-4} &	176& \num{5.1d-5} &	\num{1.3d-4} &	\num{3.1d-4} \\
    \hline
    b	    & 88 & \num{2.1d-6}	& \num{2.2d-5} & {\color{blue}\num{5.8d-5}} &	27 & {\color{blue}\num{9.1d-11}} &	{\color{blue}\num{4.3d-6}} &	{\color{blue}\num{1.8d-5}} \\
    \hline
  \end{tabular}
  \caption{The distributions of betweennesses of papers in 1999.04 and 1999.13 that share common references with the other TCs in 1999 (1999.00 to 1999.15). Four columns below `1999.04' and `1999.13' denote: the first column shows how many papers have common references with the other TCs, while the second, third, and fourth column show the lower, median, and upper quartile values of betweennesses of these papers, respectively. For example, there are 25 papers in 1999.04 that share common references with papers in 1999.03, and the betweennesses of these papers have a lower quartile value of \num{1.6d-5}, a median value of \num{4.3d-4}, and an upper quartile value of \num{8.1d-4}. Similarly, there are 254 papers in 1999.13 that share common references with papers in 1999.10, and the betweennesses of these papers have a lower quartile value of \num{3.6d-5}, a median value of \num{8.8d-5}, and an upper quartile value of \num{2.7d-4}. The bottom row `b' represent 1999.04b and 1999.13b respectively, which are papers in 1999.04 and 1999.13 have no references in common with papers in other TCs. A betweenness value in red means that it is larger than the maximum of the corresponding quartile distribution of \num{e6} random samples, and a betweenness value in blue denotes it is smaller than the minimum of the corresponding \num{e6} random samples.}\label{tab:continue_percentile}
\end{table}
However, as we can conclude from \autoref{tab:continue_percentile}, while the lower betweenness quartiles of the coupling parts of 1999.04 and 1999.13 with other TCs may be significantly larger than those of random samples of the two TCs, the highest betweenness quartiles are never significantly larger.
Therefore, at the same level of confidence that we have set for the precursors of merging between 1999.01 and 1999.02, as well as between 1999.11 and 1999.12, we have to say that there is no significant precursors for 1999.04 and 1999.13 to merge with other TCs.

What about splitting then?
A TC is likely to split into two if at least one of two parts has reduced betweenness.
We see in \autoref{tab:continue_percentile} that betweenness in the coupling parts of 1999.04 and 1999.13 are not significantly lower than those of random samples.
Therefore, we look at the non-coupling part, i.e. papers in 1999.04 and 1999.13 which have no references in common with papers in other TCs, but they may have common references with papers in the same TCs.
We call these non-coupling parts $1999.04b$ and $1999.13b$, respectively (the bottom row in \autoref{tab:continue_percentile}).
Only the top betweenness quartile of $1999.04b$ falls below that of random samples from 1999.04 in \autoref{tab:continue_percentile}.
Therefore, the early warning for a splitting event in the next year is not strong enough.
For $1999.13b$, on the other hand, all three betweenness quartiles fall below that of random samples from 1999.13, even after we have accounted for the small size of $1999.13b$.
This suggests that the probabiiity of a splitting event next year is high, but 1999.13 continued on to 2000.16, which thereafter continued to 2001 without merging or splitting.
This might be because additional conditions, like the size of TC being large, must be satisfied before a splitting can occur.

\section*{Discussion}

During the past two decades, researchers have made a lot of efforts to understand the system of science.
Many problems are solved, however the understanding of interactions between different fields is still limited.
Investigating the temporal network (BCN, CN) and their community structures, we are able to measure and quantify the complex interaction between different fields particularly in physics over time.
Naturally, we would like to have a predictive power based on this picture.
However, the correlation between network structure and evolution events is nonlinear and complex.
Therefore we turn to machine learning techniques, which have shown a great power to solve predictive problems that are hardly to be solved using traditional statistical methods.
To our knowledge, this is the first study that utilizes both machine learning and network science approaches to predict the future of science at the community level.

To be able to identify changes in TCs we needed to define time windows used for network creation and community detection.
The natural choice for bibliographical data was the usage of single years, since the publishing process may last many months.
Obviously, another granularity may be considered like multiple years, e.g. 2 or 5 years.
In our approach, i.e. both for BCN and CN, every citation has the same importance.
However, there are some other concepts like fractional counting of citations \cite{Leydesdorff2010}.
It assumes that the impact of each citation is proportionate to the number of references in the citing document.
Additionally, it can be differentiated depending on e.g. the quality of the journal.
For the CN we have calculated the similarity between groups in the consecutive time windows in two ways: (i) using the plain relative overlap measure, and (ii) using the inclusion measure based on Social Position.
The idea was to enrich evolution data with the structural information occurring between the nodes.
It turned out that both measures provided the similar labelling, but the evolution tracking with the Social Position information produced slightly better initial prediction.
Therefore, the study was continued only for the inclusion measure,see SI for more information.

We decided to analyse more in-depth only on one feature describing structural profile of TCs, namely node betweenness.
It was primary caused by the limited amount of resources and complexity of analyses.
The entire process required much human assistance and could not have been easily automated.
In our experiments, we utilized the raw, imbalanced or artificially flattened---balanced data sets.
However, depending on the study purpose, we can bias some classes we are more interested in e.g. split.
It can be achieved either by means of appropriate balancing---sampling for the learning set, or reformulating the problem into the binary question---is split expected (true) or not (false).
As of now, the betweenness analysis is still limited to several case studies, in future a more rigorous framework will be desired.
The idea of analysing science by discovery of knowledge changes is general and can be applied to all bibliographical data containing citations. We focus solely on APS journals, however, also papers indexed by PubMed, Web of Science or Google Scholar may be studied.

\section*{Methods}

The entire analytical process consists of several steps that are primary defined by the Group Evolution Prediction (GEP) framework.
First, the bibliographic coupling network (BCN) and co-citation network (CN) are extracted from the references placed in the papers from a given time window, see \autoref{fig:Fig_method_1}, and this is carried out separately for each period.
As a result, we get a time series of BCNs/CNs.
Next, paper groups called topical clusters (TCs) are extracted using the Louvain clustering methods, independently for each BCN/CN in the time series.
Each group is described by the set of predictive features.
Having TCs for consecutive periods, we were able to identify changes in TC evolution using the Group Evolution Discovery (GED) method that appropriately labels the TC changes, see below.

Independently, the features ranking and its validation were performed to find the most valuable TC measures.
Based on this ranking, a structural measure node betweenness was selected for the more in-depth studies as the early signal for splitting or merging.

\subsection*{GEP method}
The Group Evolution Prediction (GEP) method is the first generic approach for the prediction of the evolution of groups\cite{Saganowski2017}, in our case groups correspond to TCs.
The GEP process consists of six main steps: (1) time window definition, (2) temporal network creation, (3) group detection, (4) group evolution tracking, (5) evolution chain identification and feature calculation, and (6) classification using machine learning techniques.
Thanks to its adaptable character, we were able to apply it to the BCN and CN differently.
For the group (TC) detection in both networks, we applied the Louvain method\cite{Blondel2008}.
The group evolution tracking was performed with the GED method (see below), but we used different similarity measures for each network BCN and CN, see below.
The set of features describing the group at a given time windows was adjusted to our networks, as some of the features defined in the GEP method were not applicable in our case.
We also introduced some new, dedicated measures appropriate for bibliographical data, see SI for the complete list.
Finally, we applied the Auto-WEKA tool to find the best predictive model and its parameters from the wide range of all possible solutions.
The commonly known average F-measure was used as a prediction performance measure.

\subsection*{Bibliographic coupling network (BCN) and co-citation network (CN)}
\begin{figure}
  \centering
  \includegraphics[width=\linewidth]{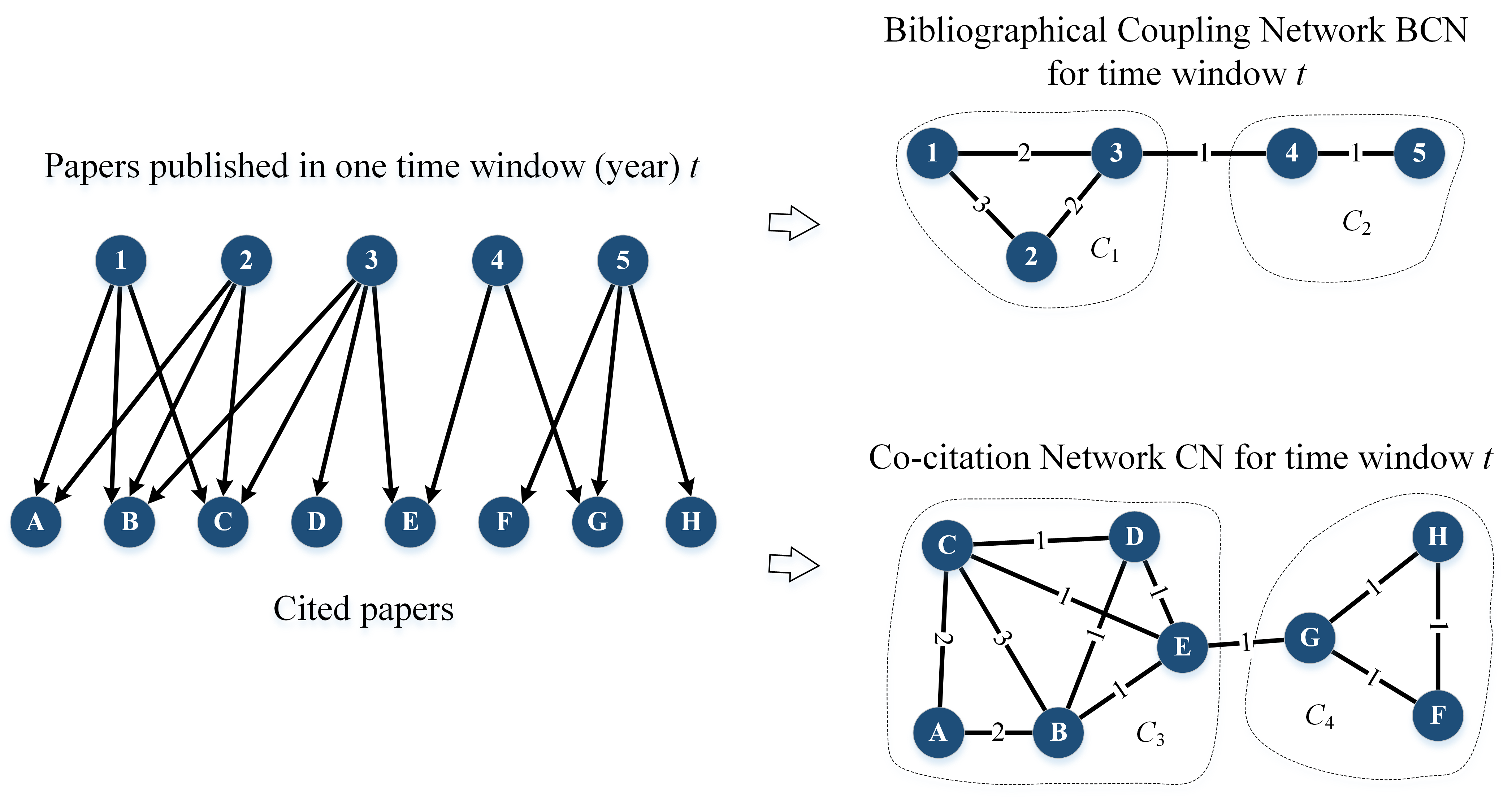}
  \caption{The process of building a Bibliographical Coupling Network (BCN) and Co-citation Network (CN) from the citation bipartite network for a given period---year $t$. Both BCN and CN are undirected and weighted; the weights denote the number of shared citations (BCN) or co-citing papers (CN). Separate topical clusters are extracted for BCN ($C_1$, $C_2$) and CN ($C_3$, $C_4$). Nodes with numbers are papers from a given period being considered and nodes with letters are their references.}\label{fig:Fig_method_1}
\end{figure}

In the BCN and CN, nodes represent papers and undirected but weighted edges denote the bibliographic coupling strengths and co-citation strengths, respectively.
That is, if two papers share $w$ common references, the BCN edge between them would have a weight of $w$.
For example, papers 1 and 2 in \autoref{fig:Fig_method_1} share three citations: A, B, and C, whereas papers 3 and 4 commonly cite only one paper---E.
On the other hand, if two papers are cited together by $w'$ papers, the edge between them in the CN receives weight $w’$.
Papers A and B are cited together by two other papers: 1 and 2, but papers B and C by three, i.e. additionally by paper 3.
Both BCN and CN are temporal networks, in which the nodes are all papers published within a specific time window (BCN) or papers cited within a given time window (CN).
We assume that the reasonable time window for bibliographical data is one year to facilitate the analysis of changes in scientific knowledge, i.e. changes in topical clusters year by year.
For the BCN, only the giant component, which in most cases occupies 99\% of the whole BCN, will be considered for the TC detection and evolution analysis.
For the CN, we do not use all papers cited in the given time window because most of them are cited only a small number of times, and thus they have little influence on the broader knowledge evolution.
Therefore, we rank all available $N$ papers ${p_1,p_2,…,p_N}$ in the descending order by the number of times they are cited in this time window (year): ${f_1,f_2,…,f_N },f_1 \geq f_2 \geq ... \geq f_N$.
Next, we choose the top $n$ papers ${p_1,p_2,…,p_n}$, that totally gathered $\frac{1}{4}$ of all citations, i.e. such that $n < N$ is the smallest integer to satisfy $\sum_{i=1}^{n}f_i \geq \frac{1}{4}\sum_{j=1}^{N}f_j$.
The data we used in this paper is the APS data set, consisting of about half a million publications between 1893 to 2013 and six million citation relations among them\cite{APS_dataset}.

\subsection*{Intimacy indices}
To analyse the evolution of TCs, we need to match them from the consecutive years.
The set of cited papers to large extent overlaps year by year, so for the CN, we can use the regular approach proposed together with the GED method, see below and Brodka \emph{et al.}\cite{Brodka2013a}.
For BCN, however, there is no overlap at all between papers published in the successive years because every paper can be published only once and in only one year.
Even if we do not have the corresponding papers in TCs from two BCNs, i.e. two years, the papers' references overlap each another.
Therefore, we can measure the similarity of their reference pools to reflect their inheritance.
For that purpose, we introduced the \emph{forward intimacy index} and \emph{backward intimacy index} in Liu \emph{et al.}\cite{Liu2017a}.
The idea behind intimacy indices is that the references related to a particular topic change gradually.
The \emph{forward intimacy index} $I^{f}_{mn}$ and the \emph{backward intimacy index} $I^{b}_{mn}$ between TCs $C_m^t$ in year $t$ and $C_n^{t+1}$ in year $t+1$ are defined in following:
\begin{equation}\label{eq:intimate_index}
  \begin{aligned}
    I^f_{mn} &= \sum_{i} \frac{N\left(R_i, \mathcal{R}_n^{t+1}\right)}{N\left(R_i, \mathcal{R}^{t+1}\right)} \frac{N\left(R_i, \mathcal{R}_m^t\right)}{L\left(\mathcal{R}_m^t\right)}, \\
    I^b_{mn} &= \sum_{i} \frac{N\left(R_i, \mathcal{R}_m^t\right)}{N\left(R_i, \mathcal{R}^{t}\right)} \frac{N\left(R_i, \mathcal{R}_n^{t+1}\right)}{L\left(\mathcal{R}_n^{t+1}\right)}.
  \end{aligned}
\end{equation}
Here the TCs at $t$ and $t+1$ are $\mathcal{C}^t = \left\{C_1^t, ...,C_m^t,..., C_u^t\right\}$ and $\mathcal{C}^{t+1} = \left\{C_1^{t+1}, ...,C_n^{t+1},..., C_v^{t+1}\right\}$,
and we denote the references cited by papers in $C^t_m$  and $C^{t+1}_n$ as {\color{black}$\mathcal{R}_m^t = \mathcal{R}(C_m^t) = \left[R_{m1}, ..., R_{mp}\right]$ and $\mathcal{R}_n^{t+1} = \mathcal{R}(C_n^{t+1}) = \left[R_{n1}, ..., R_{nq}\right]$}; and $\mathcal{R}^t = \left\{\mathcal{R}_1^t, ...,\mathcal{R}_m^t,...\right\}$.
$N(element, list)$ is the number of times $element$ occurs in $list$, and $L(list)$ is the length of $list$.
For more details and examples of intimacy indices, please refer to Liu \emph{et al.} \cite{Liu2017a}.

\subsection*{GED method}
The Group Evolution Discovery (GED) method\cite{Brodka2013a} was used for tracking group evolution for historical cases--—to learn the classifier and for testing cases to validate classification results.
The GED method makes use of the similarity between groups in the following years as well as their sizes to label one of six event types: continuing, dissolving, merging, splitting, growing, shrinking.
However, we have adapted the GED method to label only four types of events: continuing, dissolving, merging, splitting, as these are the most important to us.
The other two (growing and shrinking) are covered by continuing.
In general, the GED method allows us to use various metrics as a similarity measure between groups.
Therefore, the intimacy indices defined in \autoref{eq:intimate_index} were used for the BCN to match similar groups in the consecutive time windows.
However, the original GED inclusion measures were used for the CN.
It means that the similarity between two groups from two successive time windows is reflected by the inclusion measure, which is calculated for two scenarios: inclusion $I(C_n^t, C_m^{t+1})$ of a group $C_n^t$ from time window $t$ in another group $C_m^{t+1}$ from time window $t+1$ (forward, \autoref{eq:forward_inclusion}), and inclusion $I(C_m^{t+1}, C_n^t)$ of this second group $C_m^{t+1}$ from $t+1$ in the first group $C_n^t$ from $t$ (backward, \autoref{eq:backward_inclusion}).
The inclusion measure makes use of the Social Position $SP(p)$, which is a kind of weighted PageRank.
It denotes an importance of paper $p$ being cited among all other papers\cite{Brodka2009}.
The inclusions for CN are defined as follows:
\begin{equation}\label{eq:forward_inclusion}
  I(C_n^t, C_m^{t+1})=\overbrace{\frac{\Vert C_n^t \cap C_m^{t+1} \Vert}{\Vert C_n^t \Vert}}^{\text{group quantity}} \cdot \underbrace{\frac{\sum\limits_{p \in (C_n^t \cap C_m^{t+1})} SP(p)}{\sum\limits_{p \in (C_n^t)} SP(p)}}_{\text{group quality}} \cdot 100\%,
\end{equation}
\begin{equation}\label{eq:backward_inclusion}
  I(C_m^{t+1}, C_n^t) = \overbrace{\frac{\Vert C_m^{t+1} \cap C_n^t \Vert}{\Vert C_m^{t+1} \Vert}}^{\text{group quantity}} \cdot \underbrace{\frac{\sum\limits_{p \in (C_m^{t+1} \cap C_n^t)} SP(p)}{\sum\limits_{p \in (C_m^{t+1})} SP(p)}}_{\text{group quality}} \cdot 100\%.
\end{equation}
If both inclusions (CN) or both intimacy indices (BCN) are greater than the percentage thresholds alpha and beta (the only parameters in this method), the method labels the event continuing.
If at least one inclusion or one intimacy index exceeds one of the thresholds, the splitting and merging events is considered, the proper event is assigned depending on the number of similar groups in $t$ and $t+1$.
If both inclusions or both intimacy indexes are below the thresholds, i.e. the group has no corresponding group in the next time window, the dissolving event is assigned.

\subsection*{Feature Ranking}
Rankings of the most prominent features was obtained by repeating the feature selection 1000 times using a basic evolutionary algorithm \cite{Yang1998}, as proposed in Saganowski \emph{et al.}\cite{Saganowski2017}.
The rankings were received for the 30 years span (1981-2010).
Next, only top 10 features were selected to described TCs in two additional years (2010-2012) and predict TC evolution.
The results revealed the superiority of feature selection compared to the raw approach with all features engagement.


\begin{thebibliography}{10}
\expandafter\ifx\csname url\endcsname\relax
  \def\url#1{\texttt{#1}}\fi
\expandafter\ifx\csname urlprefix\endcsname\relax\def\urlprefix{URL }\fi
\expandafter\ifx\csname doiprefix\endcsname\relax\def\doiprefix{DOI }\fi
\providecommand{\bibinfo}[2]{#2}
\providecommand{\eprint}[1]{\href{https://dx.doi.org/#1}{#1}}

\bibitem{Chen2010}
\bibinfo{author}{Chen, P.} \& \bibinfo{author}{Redner, S.}
\newblock \bibinfo{journal}{\bibinfo{title}{{Community structure of the
  physical review citation network}}}.
\newblock {\emph{\JournalTitle{Journal of Informetrics}}}
  \textbf{\bibinfo{volume}{4}}, \bibinfo{pages}{278--290}
  (\bibinfo{year}{2010}).
\newblock \doiprefix \eprint{10.1016/j.joi.2010.01.001}.

\bibitem{Rosvall2010}
\bibinfo{author}{Rosvall, M.} \& \bibinfo{author}{Bergstrom, C.~T.}
\newblock \bibinfo{journal}{\bibinfo{title}{{Mapping change in large
  networks}}}.
\newblock {\emph{\JournalTitle{PLoS ONE}}} \textbf{\bibinfo{volume}{5}}
  (\bibinfo{year}{2010}).
\newblock \doiprefix \eprint{10.1371/journal.pone.0008694}.

\bibitem{Liu2017a}
\bibinfo{author}{Liu, W.}, \bibinfo{author}{Nanetti, A.} \&
  \bibinfo{author}{Cheong, S.~A.}
\newblock \bibinfo{journal}{\bibinfo{title}{{Knowledge evolution in physics
  research: An analysis of bibliographic coupling networks}}}.
\newblock {\emph{\JournalTitle{PLoS ONE}}} \textbf{\bibinfo{volume}{12}},
  \bibinfo{pages}{1--19} (\bibinfo{year}{2017}).
\newblock \doiprefix \eprint{10.1371/journal.pone.0184821}.

\bibitem{Zeng2017}
\bibinfo{author}{Zeng, A.} \emph{et~al.}
\newblock \bibinfo{journal}{\bibinfo{title}{{The science of science: From the
  perspective of complex systems}}}.
\newblock {\emph{\JournalTitle{Physics Reports}}}
  \textbf{\bibinfo{volume}{714-715}}, \bibinfo{pages}{1--73}
  (\bibinfo{year}{2017}).
\newblock \doiprefix \eprint{10.1016/j.physrep.2017.10.001}.

\bibitem{Fortunato2018}
\bibinfo{author}{Fortunato, S.} \emph{et~al.}
\newblock \bibinfo{journal}{\bibinfo{title}{{Science of science}}}.
\newblock {\emph{\JournalTitle{Science}}} \textbf{\bibinfo{volume}{359}},
  \bibinfo{pages}{eaao0185} (\bibinfo{year}{2018}).
\newblock \doiprefix \eprint{10.1126/science.aao0185}.

\bibitem{Hicks2015}
\bibinfo{author}{Hicks, D.}, \bibinfo{author}{Wouters, P.},
  \bibinfo{author}{Waltman, L.}, \bibinfo{author}{{De Rijcke}, S.} \&
  \bibinfo{author}{Rafols, I.}
\newblock \bibinfo{journal}{\bibinfo{title}{{Bibliometrics: The Leiden
  Manifesto for research metrics}}}.
\newblock {\emph{\JournalTitle{Nature}}} \textbf{\bibinfo{volume}{520}},
  \bibinfo{pages}{429--431} (\bibinfo{year}{2015}).
\newblock \doiprefix \eprint{10.1038/520429a}.

\bibitem{Radicchi2008}
\bibinfo{author}{Radicchi, F.}, \bibinfo{author}{Fortunato, S.} \&
  \bibinfo{author}{Castellano, C.}
\newblock \bibinfo{journal}{\bibinfo{title}{{Universality of citation
  distributions: Toward an objective measure of scientific impact}}}.
\newblock {\emph{\JournalTitle{Proceedings of the National Academy of
  Sciences}}} \textbf{\bibinfo{volume}{105}}, \bibinfo{pages}{17268--17272}
  (\bibinfo{year}{2008}).
\newblock \doiprefix \eprint{10.1073/pnas.0806977105}.

\bibitem{Wang2013}
\bibinfo{author}{Wang, D.}, \bibinfo{author}{Song, C.} \&
  \bibinfo{author}{Barab{\'{a}}si, A.-L.}
\newblock \bibinfo{journal}{\bibinfo{title}{{Quantifying Long-Term Scientific
  Impact}}}.
\newblock {\emph{\JournalTitle{Science}}} \textbf{\bibinfo{volume}{342}},
  \bibinfo{pages}{127--132} (\bibinfo{year}{2013}).
\newblock \doiprefix \eprint{10.1126/science.1237825}.

\bibitem{Ke2015}
\bibinfo{author}{Ke, Q.}, \bibinfo{author}{Ferrara, E.},
  \bibinfo{author}{Radicchi, F.} \& \bibinfo{author}{Flammini, A.}
\newblock \bibinfo{journal}{\bibinfo{title}{{Defining and identifying Sleeping
  Beauties in science}}}.
\newblock {\emph{\JournalTitle{Proceedings of the National Academy of
  Sciences}}} \textbf{\bibinfo{volume}{112}}, \bibinfo{pages}{7426--7431}
  (\bibinfo{year}{2015}).
\newblock \doiprefix \eprint{10.1073/pnas.1424329112}.

\bibitem{Small1999}
\bibinfo{author}{Small, H.}
\newblock \bibinfo{journal}{\bibinfo{title}{{Visualizing science by citation
  mapping}}}.
\newblock {\emph{\JournalTitle{Journal of the American Society for Information
  Science}}} \textbf{\bibinfo{volume}{50}}, \bibinfo{pages}{799--813}
  (\bibinfo{year}{1999}).
\newblock \doiprefix \eprint{10.1002/(SICI)1097-4571(1999)50:9<799::AID-ASI9>3.0.CO;2-G}.

\bibitem{Boyack2005}
\bibinfo{author}{Boyack, K.~W.}, \bibinfo{author}{Klavans, R.} \&
  \bibinfo{author}{B{\"{o}}rner, K.}
\newblock \bibinfo{journal}{\bibinfo{title}{{Mapping the backbone of
  science}}}.
\newblock {\emph{\JournalTitle{Scientometrics}}} \textbf{\bibinfo{volume}{64}},
  \bibinfo{pages}{351--374} (\bibinfo{year}{2005}).
\newblock \doiprefix \eprint{10.1007/s11192-005-0255-6}.

\bibitem{Bollen2009}
\bibinfo{author}{Bollen, J.} \emph{et~al.}
\newblock \bibinfo{journal}{\bibinfo{title}{{Clickstream Data Yields
  High-Resolution Maps of Science}}}.
\newblock {\emph{\JournalTitle{PLoS ONE}}} \textbf{\bibinfo{volume}{4}},
  \bibinfo{pages}{e4803} (\bibinfo{year}{2009}).
\newblock \doiprefix \eprint{10.1371/journal.pone.0004803}.

\bibitem{Palla2007}
\bibinfo{author}{Palla, G.}, \bibinfo{author}{Barab{\'{a}}si, A.-L.} \&
  \bibinfo{author}{Vicsek, T.}
\newblock \bibinfo{journal}{\bibinfo{title}{{Quantifying social group
  evolution}}}.
\newblock {\emph{\JournalTitle{Nature}}} \textbf{\bibinfo{volume}{446}},
  \bibinfo{pages}{664--667} (\bibinfo{year}{2007}).
\newblock \doiprefix \eprint{10.1038/nature05670}.

\bibitem{Carrasquilla2017}
\bibinfo{author}{Carrasquilla, J.} \& \bibinfo{author}{Melko, R.~G.}
\newblock \bibinfo{journal}{\bibinfo{title}{{Machine learning phases of
  matter}}}.
\newblock {\emph{\JournalTitle{Nature Physics}}} \textbf{\bibinfo{volume}{13}},
  \bibinfo{pages}{431--434} (\bibinfo{year}{2017}).
\newblock \doiprefix \eprint{10.1038/nphys4035}.

\bibitem{Ahneman2018}
\bibinfo{author}{Ahneman, D.~T.}, \bibinfo{author}{Estrada, J.~G.},
  \bibinfo{author}{Lin, S.}, \bibinfo{author}{Dreher, S.~D.} \&
  \bibinfo{author}{Doyle, A.~G.}
\newblock \bibinfo{journal}{\bibinfo{title}{{Predicting reaction performance in
  C–N cross-coupling using machine learning}}}.
\newblock {\emph{\JournalTitle{Science}}} \textbf{\bibinfo{volume}{360}},
  \bibinfo{pages}{186--190} (\bibinfo{year}{2018}).
\newblock \doiprefix \eprint{10.1126/science.aar5169}.

\bibitem{Saganowski2017}
\bibinfo{author}{Saganowski, S.}, \bibinfo{author}{Br{\'{o}}dka, P.},
  \bibinfo{author}{Koziarski, M.} \& \bibinfo{author}{Kazienko, P.}
\newblock \bibinfo{journal}{\bibinfo{title}{{Analysis of group evolution
  prediction in complex networks}}}.
\newblock {\emph{\JournalTitle{arXiv preprint}}}  (\bibinfo{year}{2017}).
\newblock \eprint{1711.01867}.

\bibitem{Saganowski2015}
\bibinfo{author}{Saganowski, S.} \emph{et~al.}
\newblock \bibinfo{journal}{\bibinfo{title}{{Predicting Community Evolution in
  Social Networks}}}.
\newblock {\emph{\JournalTitle{Entropy}}} \textbf{\bibinfo{volume}{17}},
  \bibinfo{pages}{3053--3096} (\bibinfo{year}{2015}).
\newblock \doiprefix \eprint{10.3390/e17053053}.

\bibitem{Ilhan2016}
\bibinfo{author}{İlhan, N.} \&
  \bibinfo{author}{{\"{O}}ğ{\"{u}}d{\"{u}}c{\"{u}}, u.~G.}
\newblock \bibinfo{journal}{\bibinfo{title}{{Feature identification for
  predicting community evolution in dynamic social networks}}}.
\newblock {\emph{\JournalTitle{Engineering Applications of Artificial
  Intelligence}}} \textbf{\bibinfo{volume}{55}}, \bibinfo{pages}{202--218}
  (\bibinfo{year}{2016}).
\newblock \doiprefix \eprint{10.1016/j.engappai.2016.06.003}.

\bibitem{Pavlopoulou2017}
\bibinfo{author}{Pavlopoulou, M. E.~G.}, \bibinfo{author}{Tzortzis, G.},
  \bibinfo{author}{Vogiatzis, D.} \& \bibinfo{author}{Paliouras, G.}
\newblock \bibinfo{title}{{Predicting the evolution of communities in social
  networks using structural and temporal features}}.
\newblock In \emph{\bibinfo{booktitle}{2017 12th International Workshop on
  Semantic and Social Media Adaptation and Personalization (SMAP)}},
  \bibinfo{pages}{40--45} (\bibinfo{publisher}{IEEE}, \bibinfo{year}{2017}).
\newblock \doiprefix \eprint{10.1109/SMAP.2017.8022665}.

\bibitem{Brodka2013a}
\bibinfo{author}{Br{\'{o}}dka, P.}, \bibinfo{author}{Saganowski, S.} \&
  \bibinfo{author}{Kazienko, P.}
\newblock \bibinfo{journal}{\bibinfo{title}{{GED: the method for group
  evolution discovery in social networks}}}.
\newblock {\emph{\JournalTitle{Social Network Analysis and Mining}}}
  \textbf{\bibinfo{volume}{3}}, \bibinfo{pages}{1--14} (\bibinfo{year}{2013}).
\newblock \doiprefix \eprint{10.1007/s13278-012-0058-8}.

\bibitem{Tajeuna2015}
\bibinfo{author}{Tajeuna, E.~G.}, \bibinfo{author}{Bouguessa, M.} \&
  \bibinfo{author}{Wang, S.}
\newblock \bibinfo{title}{{Tracking the evolution of community structures in
  time-evolving social networks}}.
\newblock In \emph{\bibinfo{booktitle}{2015 IEEE International Conference on
  Data Science and Advanced Analytics (DSAA)}}, \bibinfo{pages}{1--10}
  (\bibinfo{publisher}{IEEE}, \bibinfo{year}{2015}).
\newblock \doiprefix \eprint{10.1109/DSAA.2015.7344876}.

\bibitem{Saganowski2017a}
\bibinfo{author}{Br{\'{o}}dka, P.}, \bibinfo{author}{Saganowski, S.} \&
  \bibinfo{author}{Kazienko, P.}
\newblock \bibinfo{title}{{Community Evolution}}.
\newblock In \emph{\bibinfo{booktitle}{Encyclopedia of Social Network Analysis
  and Mining}}, \bibinfo{pages}{220--232} (\bibinfo{publisher}{Springer New
  York}, \bibinfo{address}{New York, NY}, \bibinfo{year}{2014}).
\newblock \doiprefix \eprint{10.1007/978-1-4614-6170-8\_223}.

\bibitem{Brodka2009}
\bibinfo{author}{Brodka, P.}, \bibinfo{author}{Musial, K.} \&
  \bibinfo{author}{Kazienko, P.}
\newblock \bibinfo{title}{{A Performance of Centrality Calculation in Social
  Networks}}.
\newblock In \emph{\bibinfo{booktitle}{2009 International Conference on
  Computational Aspects of Social Networks}}, \bibinfo{pages}{24--31}
  (\bibinfo{publisher}{IEEE}, \bibinfo{year}{2009}).
\newblock \doiprefix \eprint{10.1109/CASoN.2009.20}.

\bibitem{Popper1999}
\bibinfo{author}{Popper, K.~R.}
\newblock \emph{\bibinfo{title}{{All life is problem solving}}}
  (\bibinfo{publisher}{Routledge}, \bibinfo{year}{2013}).

\bibitem{Kotthoff2017}
\bibinfo{author}{Kotthoff, L.}, \bibinfo{author}{Thornton, C.},
  \bibinfo{author}{Hoos, H.~H.}, \bibinfo{author}{Hutter, F.} \&
  \bibinfo{author}{Leyton-Brown, K.}
\newblock \bibinfo{journal}{\bibinfo{title}{{Auto-WEKA 2.0: Automatic model
  selection and hyperparameter optimization in WEKA}}}.
\newblock {\emph{\JournalTitle{Journal of Machine Learning Research}}}
  \textbf{\bibinfo{volume}{17}}, \bibinfo{pages}{1--5} (\bibinfo{year}{2016}).

\bibitem{Platt1999}
\bibinfo{author}{Platt, J.~C.}
\newblock \bibinfo{title}{{Fast training of support vector machines using
  sequential minimal optimization}}.
\newblock In \emph{\bibinfo{booktitle}{Advances in kernel methods}},
  \bibinfo{pages}{185----208} (\bibinfo{publisher}{IEEE},
  \bibinfo{year}{1999}).

\bibitem{Christopher1997}
\bibinfo{author}{Christopher, A.}, \bibinfo{author}{Andrew, M.} \&
  \bibinfo{author}{Stefan, S.}
\newblock \bibinfo{journal}{\bibinfo{title}{{Locally Weighted Learning}}}.
\newblock {\emph{\JournalTitle{Artif Intell Rev}}}
  \textbf{\bibinfo{volume}{11}}, \bibinfo{pages}{11--73}
  (\bibinfo{year}{1997}).

\bibitem{Frank1998}
\bibinfo{author}{Frank, E.} \& \bibinfo{author}{Witten, I.~H.}
\newblock \bibinfo{title}{{Generating accurate rule sets without global
  optimization}}.
\newblock In \emph{\bibinfo{booktitle}{Proceedings of the Fifteenth
  International Conference on Machine Learning}}, \bibinfo{pages}{144----151}
  (\bibinfo{publisher}{Morgan Kaufmann Publishers Inc.}, \bibinfo{year}{1998}).

\bibitem{Yang1998}
\bibinfo{author}{Yang, J.} \& \bibinfo{author}{Honavar, V.}
\newblock \bibinfo{title}{{Feature Subset Selection Using a Genetic
  Algorithm}}.
\newblock In \emph{\bibinfo{booktitle}{Feature Extraction, Construction and
  Selection}}, \bibinfo{pages}{117--136} (\bibinfo{publisher}{Springer US},
  \bibinfo{address}{Boston, MA}, \bibinfo{year}{1998}).
\newblock \doiprefix \eprint{10.1007/978-1-4615-5725-8\_8}.

\bibitem{Leydesdorff2010}
\bibinfo{author}{Leydesdorff, L.} \& \bibinfo{author}{Opthof, T.}
\newblock \bibinfo{journal}{\bibinfo{title}{{Scopus's source normalized impact
  per paper (SNIP) versus a journal impact factor based on fractional counting
  of citations}}}.
\newblock {\emph{\JournalTitle{Journal of the American Society for Information
  Science and Technology}}} \textbf{\bibinfo{volume}{61}},
  \bibinfo{pages}{2365--2369} (\bibinfo{year}{2010}).
\newblock \doiprefix \eprint{10.1002/asi.21371}.

\bibitem{Blondel2008}
\bibinfo{author}{Blondel, V.~D.}, \bibinfo{author}{Guillaume, J.-L.},
  \bibinfo{author}{Lambiotte, R.} \& \bibinfo{author}{Lefebvre, E.}
\newblock \bibinfo{journal}{\bibinfo{title}{{Fast unfolding of communities in
  large networks}}}.
\newblock {\emph{\JournalTitle{Journal of Statistical Mechanics: Theory and
  Experiment}}} \textbf{\bibinfo{volume}{2008}}, \bibinfo{pages}{P10008}
  (\bibinfo{year}{2008}).
\newblock \doiprefix \eprint{10.1088/1742-5468/2008/10/P10008}.

\bibitem{APS_dataset}
\bibinfo{title}{APS Data Sets for Research}.
\newblock \bibinfo{howpublished}{\url{http://journals.aps.org/datasets}}.

\bibitem{Page1999}
\bibinfo{author}{Page, L.}, \bibinfo{author}{Brin, S.},
  \bibinfo{author}{Motwani, R.} \& \bibinfo{author}{Winograd, T.}
\newblock \bibinfo{title}{{The PageRank Citation Ranking: Bringing Order to the
  Web}}.
\newblock \bibinfo{type}{Tech. Rep.} (\bibinfo{year}{1999}).

\bibitem{Saganowski2012}
\bibinfo{author}{Saganowski, S.}, \bibinfo{author}{Br{\'{o}}dka, P.} \&
  \bibinfo{author}{Kazienko, P.}
\newblock \bibinfo{journal}{\bibinfo{title}{{Influence of the User Importance
  Measure on the Group Evolution Discovery}}}.
\newblock {\emph{\JournalTitle{Foundations of Computing and Decision
  Sciences}}} \textbf{\bibinfo{volume}{37}} (\bibinfo{year}{2012}).
\newblock \doiprefix \eprint{10.2478/v10209-011-0017-6}.

\bibitem{Freeman1978}
\bibinfo{author}{Freeman, L.~C.}
\newblock \bibinfo{journal}{\bibinfo{title}{{Centrality in social networks
  conceptual clarification}}}.
\newblock {\emph{\JournalTitle{Social Networks}}} \textbf{\bibinfo{volume}{1}},
  \bibinfo{pages}{215--239} (\bibinfo{year}{1978}).
\newblock \doiprefix \eprint{10.1016/0378-8733(78)90021-7}.

\bibitem{Bonacich1972}
\bibinfo{author}{Bonacich, P.}
\newblock \bibinfo{journal}{\bibinfo{title}{{Factoring and weighting approaches
  to status scores and clique identification}}}.
\newblock {\emph{\JournalTitle{The Journal of Mathematical Sociology}}}
  \textbf{\bibinfo{volume}{2}}, \bibinfo{pages}{113--120}
  (\bibinfo{year}{1972}).
\newblock \doiprefix \eprint{10.1080/0022250X.1972.9989806}.

\bibitem{Harary1969}
\bibinfo{author}{Harary, F.}
\newblock \emph{\bibinfo{title}{{Graph theory}}}
  (\bibinfo{publisher}{Addison-Wesley, Reading, MA}, \bibinfo{year}{1969}).

\bibitem{Wasserman1994}
\bibinfo{author}{Wasserman, S.} \& \bibinfo{author}{Faust, K.}
\newblock \emph{\bibinfo{title}{{Social Network Analysis: Methods and
  Applications}}} (\bibinfo{publisher}{Cambridge University Press},
  \bibinfo{address}{Cambridge}, \bibinfo{year}{1994}).
\newblock \doiprefix \eprint{10.1017/CBO9780511815478}.

\bibitem{White2001}
\bibinfo{author}{White, D.~R.} \& \bibinfo{author}{Harary, F.}
\newblock \bibinfo{journal}{\bibinfo{title}{{The Cohesiveness of Blocks In
  Social Networks: Node Connectivity and Conditional Density}}}.
\newblock {\emph{\JournalTitle{Sociological Methodology}}}
  \textbf{\bibinfo{volume}{31}}, \bibinfo{pages}{305--359}
  (\bibinfo{year}{2001}).
\newblock \doiprefix \eprint{10.1111/0081-1750.00098}.

\bibitem{Newman2010}
\bibinfo{author}{Newman, M.}
\newblock \emph{\bibinfo{title}{{Networks: An Introduction}}}
  (\bibinfo{publisher}{Oxford University Press}, \bibinfo{year}{2010}).
\newblock \doiprefix \eprint{10.1093/acprof:oso/9780199206650.001.0001}.

\end{thebibliography}

\section*{Acknowledgements}
W.L. thanks the Nanyang Technological University for supporting him through a research scholarship.
S.A.C. acknowledges support from the Singapore Ministry of Education Academic Research Fund Tier 2, under grant number MOE2017-T2-2-075.
S.S. and P.K. received partial supports from the National Science Centre, Poland, the project no. 2016/21/B/ST6/01463, from the European Union’s Marie Skłodowska-Curie Program under grant agreement no. 691152, and from the Polish Ministry of Science and Higher Education under grant agreement no. 3628/H2020/2016/2.

\section*{Author contributions statement}

W.L., S.S., S.A.C. and P.K. conceived the study. W.L., S.S. and S.A.C. designed and performed the research. W.L., S.S. and S.A.C. wrote the manuscript. W.L., S.S., S.A.C. and P.K. reviewed and approved the manuscript.

\section*{Additional information}

\textbf{Competing financial interests:} The authors declare no competing financial interests.


\pagebreak
\clearpage
\begin{center}
\textbf{\LARGE Supplementary Information}

\end{center}
\setcounter{equation}{0}
\setcounter{figure}{0}
\setcounter{table}{0}
\makeatletter
\renewcommand{\theequation}{S\arabic{equation}}
\renewcommand{\thefigure}{S\arabic{figure}}
\renewcommand{\thetable}{S\arabic{table}}

\section*{The alpha and beta thresholds for event labelling}
Two groups in the consecutive time windows are considered similar if at least one of their inclusion measures is greater than alpha or beta parameters.
In other words, the alpha and beta parameters are thresholds which have to be satisfied to assign an event between two groups.
The theoretical range of values for alpha and beta is between 0\% and 100\%.
However, the most common values are selected from the range from 30\% to 70\%, depending on the density of the network and node’s fluctuation year by year.
In general, the selection of parameters should reflect the needs of researchers.
For example, one may choose very high value (e.g. 80\%) in order to preserve only very similar groups.
In another case, it might be necessary to set very low value, e.g. 10\% if the network is sparse or the fluctuation is high.
In our study, we ran the GED method with alpha and beta parameters varying from 5\% to 100\%, to see how the number of events varies.
Our goal was to have at least one event assigned to each TC.
As the splitting and merging events involve several groups, we aimed to have on average slightly more than one event per TC.
With this assumption, we selected 30\% for both alpha and beta parameters in case of BCN, and 10\% for alpha and beta parameters in case of CN.
This values produced in total 479 events per 430 groups for BCN, and 492 events per 457 groups for CN.

\section*{Correlation between overlap measure and inclusion measure}
\begin{figure*}[h!]
  \centering
  \begin{subfigure}[t]{0.45\textwidth}
    \centering
    \includegraphics[width=\textwidth]{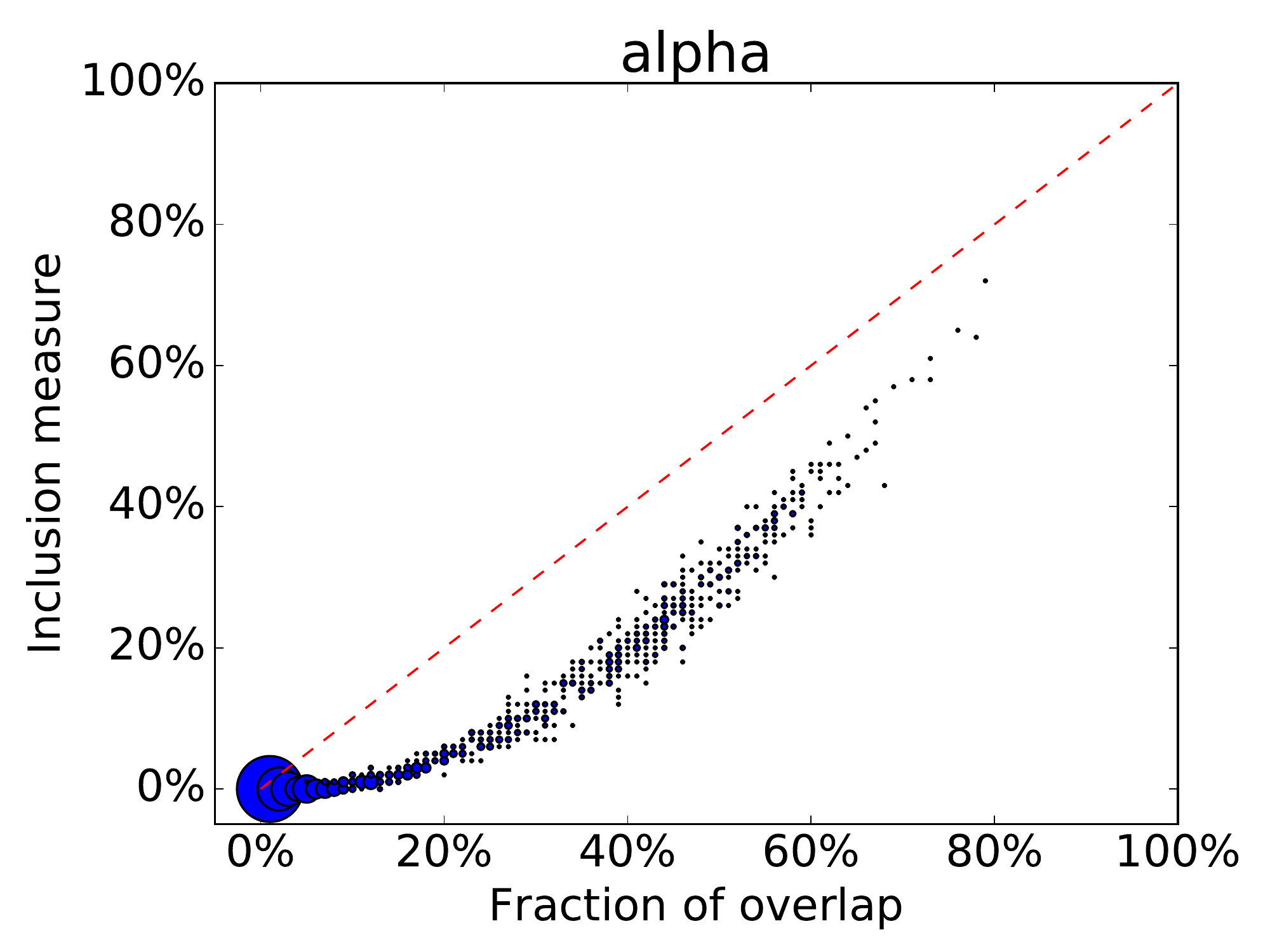}
    \caption{}
  \end{subfigure}%
  ~
  \begin{subfigure}[t]{0.45\textwidth}
    \centering
    \includegraphics[width=\textwidth]{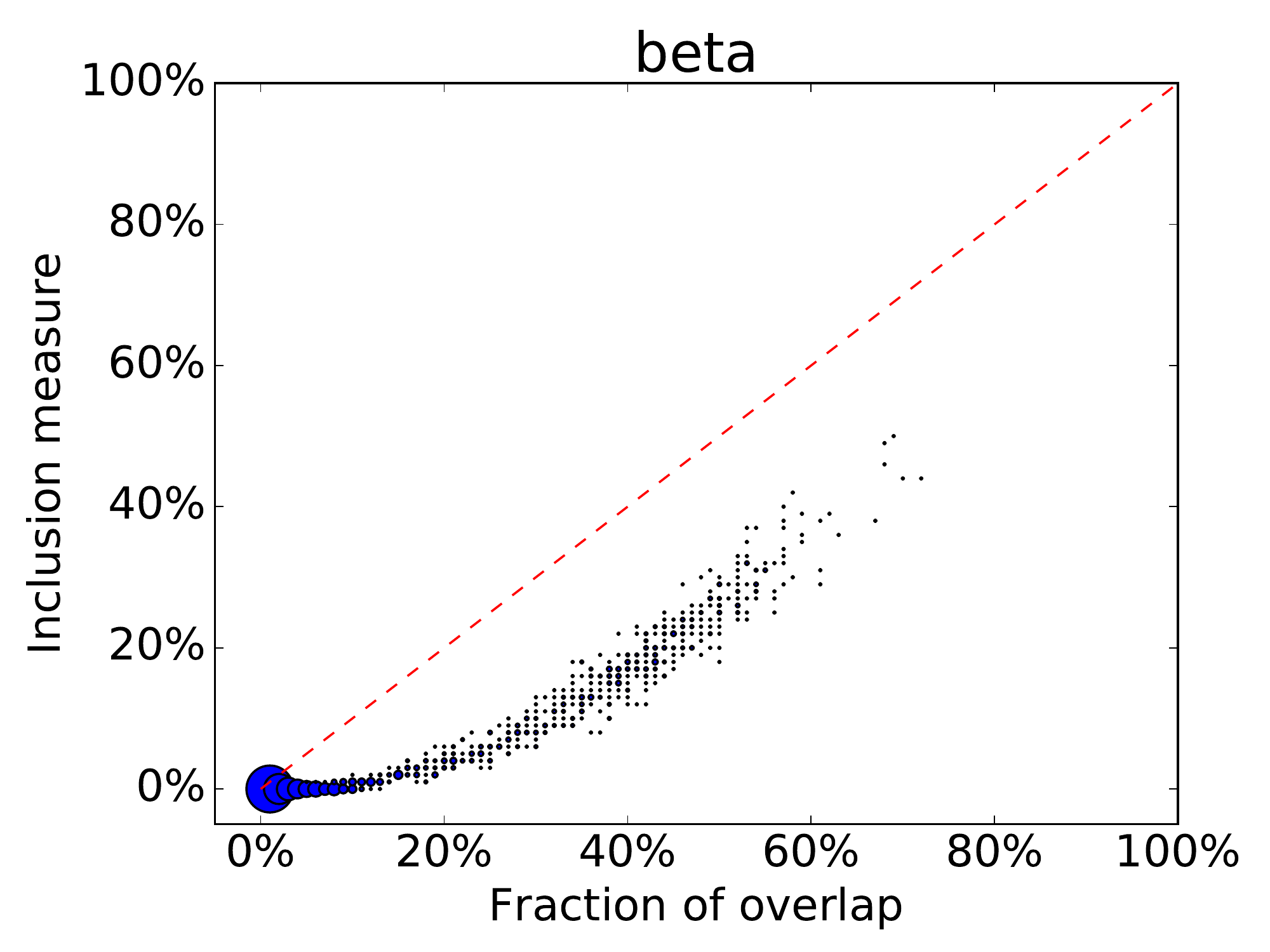}
    \caption{}
  \end{subfigure}
  \caption{The scatter plots for simple overlap measure and inclusion measure for CNs between 1981 to 2010. The left panel (A) is for the alpha parameter, i.e. how the groups in $t$ are close to groups in $t+1$. The right panel (B) is for beta parameter, i.e. how the groups in $t+1$ are close to groups in $t$. The sizes of circles are proportional to the number of instances. The red dash lines are $y = x$ for reference only.}\label{fig:FigS2}
\end{figure*}

For BCN, we use the forward and backward intimacy indices to measure the closeness between TCs in consecutive time windows (years).
For CN, we considered two types of measure: (i) a simple overlap measure of two groups (the relative fraction of common members), and (ii) an overlap of two groups enriched with the information about the importance of the common members.
The latter is suggested by the GED method authors, who named their similarity measure the inclusion measure.
One way to evaluate the importance of TC members is to use node centrality measures to rank them within the group.
In our work, we are using the Social Position measure\cite{Brodka2009} (as suggested in the GED method), an idea based on the PageRank algorithm\cite{Page1999}.
Saganowski \emph{et al.}\cite{Saganowski2012} found that using a richer similarity measure allows us to track group evolution more reliably.
To better understand the difference between the simple overlap measure and the inclusion measure we compared values obtained with both measures in \autoref{fig:FigS2}.
It turned out that the inclusion measure is on average 20\% lower than the simple overlap measure, and the corresponding values, i.e. 30\% for the simple overlap and 10\% for the inclusion measure, produce roughly the same number of the evolution events.
However, the more complex version of the similarity measure (i.e. the inclusion measure), provided slightly better initial prediction results. Therefore, we finally utilized the inclusion measure in our calculations for CN.

\section*{Alluvial diagram for CN}
Like the bibliographic coupling network (BCN), the co-citation network (CN) can also be visualized in the form of the alluvial diagram.
The groups in a CN represent the papers from the past that are coherent and related to a certain topic that stimulates the present research lines.

\begin{figure}[h!]
  \centering
  \includegraphics[angle= 90, origin=c, width=0.55\linewidth]{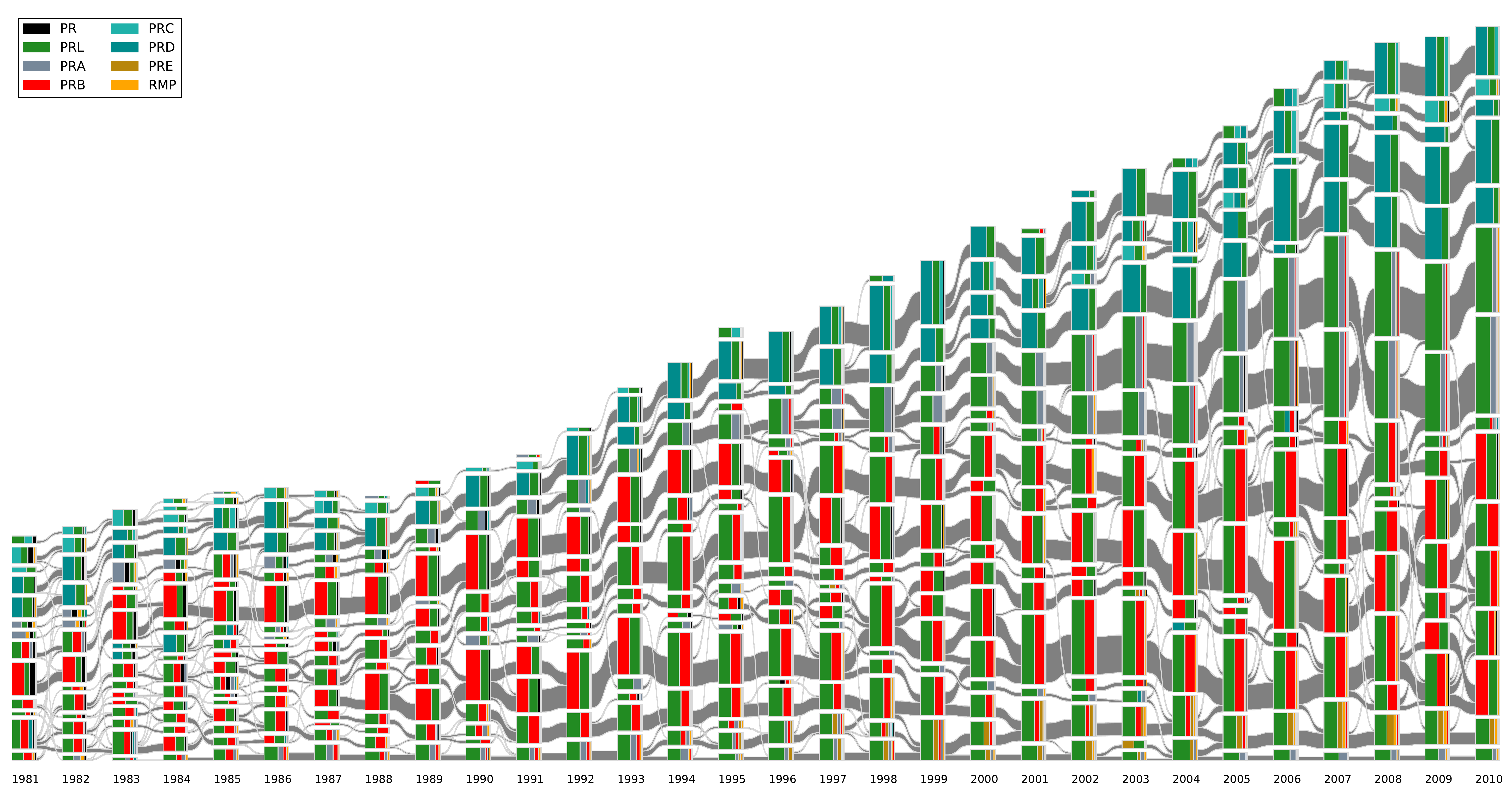}
  \caption{The alluvial diagram of APS papers’ references (CNs) from 1981 to 2010. Each block in a column represents a TC extracted from the CN. The height of the block is proportional to the number of papers in the TC. For clarity, only TCs comprising more than 1\% of all nodes in CNs are shown. TCs in successive years are connected by streams whose widths at the left and right ends are proportional to the relative overlap percentage. The colours inside a TC represent the relative contributions from different journals.}\label{fig:FigS1}
\end{figure}

\section*{The list of features used in the study}

As we mentioned in \textbf{Future events prediction}, each observation contained 77 features (preselected from the initial 100).
The full list of 100 features are showed in \autoref{tab:TabS1}.
Many features in this list are proposed for directed social network, therefore are inappropriate for our undirected BCN and CN.
The symbol $+$ indicates this feature was used in BCN prediction, while the symbol $\ast$ indicates this feature was used in CN prediction.

\begin{center}
\begin{longtable}{|c|c|p{0.42\textwidth}|}
    \hline
    \multicolumn{1}{|c|}{\textbf{Features group}} & \multicolumn{1}{c|}{\textbf{Feature name}} & \multicolumn{1}{c|}{\textbf{Feature description}} \\ \hline
    \endfirsthead

    \multicolumn{3}{c}%
    {{\bfseries \tablename\ \thetable{} -- continued from previous page}} \\
    \hline \multicolumn{1}{|c|}{\textbf{Features group}} &
    \multicolumn{1}{c|}{\textbf{Feature name}} &
    \multicolumn{1}{c|}{\textbf{Feature description}} \\ \hline
    \endhead

    \hline \multicolumn{3}{|l|}{{Continued on next page}} \\ \hline
    \endfoot

    \hline \hline
    \endlastfoot

    \multirow{53}{*}{Members/microscopic} & sum\_group\_degree\_in & The sum of indegree\cite{Freeman1978} of nodes belonging to the community calculated within the community. Indegree is a node measure defining the number of connections directed to the node \\
    \cline{2-3}
     & avg\_group\_degree\_in & The average value of indegree of nodes belonging to the community calculated within the community \\
     \cline{2-3}
     & min\_group\_degree\_in & The minimum value of indegree of nodes belonging to the community calculated within the community \\
     \cline{2-3}
     & max\_group\_degree\_in & The maximum value of indegree of nodes belonging to the community calculated within the community \\
     \cline{2-3}
     & sum\_group\_degree\_out & The sum of outdegree\cite{Freeman1978} of nodes belonging to the community calculated within the community. Outdegree is a node measure determining the number of connections outgoing from the node \\
     \cline{2-3}
     & avg\_group\_degree\_out & The average value of outdegree of nodes belonging to the community calculated within the community \\
     \cline{2-3}
     & min\_group\_degree\_out & The minimum value of outdegree of nodes belonging to the community calculated within the community \\
     \cline{2-3}
     & max\_group\_degree\_out & The maximum value of outdegree of nodes belonging to the community calculated within the community \\
     \cline{2-3}
     & sum\_group\_degree\_total$+\ast$ & The sum of total degree of nodes belonging to the community calculated within the community. Total degree is the sum of indegree and outdegree \\
     \cline{2-3}
     & avg\_group\_degree\_total$+\ast$ & The average value of total degree of nodes belonging to the community calculated within the community \\
     \cline{2-3}
     & min\_group\_degree\_total$+\ast$ & The minimum value of total degree of nodes belonging to the community calculated within the community \\
     \cline{2-3}
     & max\_group\_degree\_total$+\ast$ & The maximum value of total degree of nodes belonging to the community calculated within the community \\
     \cline{2-3}
     & sum\_group\_betweenness$+\ast$ & The sum of betweenness\cite{Freeman1978} of nodes belonging to the community calculated within the community. Betweenness is a node measure describing the number of the shortest paths from all nodes to all others that pass through that node \\
     \cline{2-3}
     & avg\_group\_betweenness$+\ast$ & The average value of betweenness of nodes belonging to the community calculated within the community \\
     \cline{2-3}
     & min\_group\_betweenness$+\ast$ & The minimum value of betweenness of nodes belonging to the community calculated within the community \\
     \cline{2-3}
     & max\_group\_betweenness$+\ast$ & The maximum value of betweenness of nodes belonging to the community calculated within the community \\
     \cline{2-3}
     & sum\_group\_closeness$+\ast$ & The sum of closeness\cite{Freeman1978} of nodes belonging to the community calculated within the community. Closeness is a node measure defined as the inverse of the farness, which in turn, is the sum of distances to all other nodes \\
     \cline{2-3}
     & avg\_group\_closeness$+\ast$ & The average value of closeness of nodes belonging to the community calculated within the community \\
     \cline{2-3}
     & min\_group\_closeness$+\ast$ & The minimum value of c of nodes belonging to the community calculated within the community \\
     \cline{2-3}
     & max\_group\_closeness$+\ast$ & The maximum value of closeness of nodes belonging to the community calculated within the community \\
     \cline{2-3}
     & sum\_group\_eigenvector$+\ast$ & The sum of eigenvector\cite{Bonacich1972} of nodes belonging to the community calculated within the community. Eigenvector is a node measure indicating the influence of a node in the network \\
     \cline{2-3}
     & avg\_group\_eigenvector$+\ast$ & The average value of eigenvector of nodes belonging to the community calculated within the community \\
     \cline{2-3}
     & min\_group\_eigenvector$+\ast$ & The minimum value of eigenvector of nodes belonging to the community calculated within the community \\
     \cline{2-3}
     & max\_group\_eigenvector$+\ast$ & The maximum value of eigenvector of nodes belonging to the community calculated within the community \\
     \cline{2-3}
     & avg\_group\_eccentricity$+\ast$ & The average value of eccentricity\cite{Harary1969} of nodes belonging to the community calculated within the community. Eccentricity of a node is its shortest path distance from the farthest other node in the graph \\
     \cline{2-3}
     & min\_group\_eccentricity$+\ast$ & The minimum value of eccentricity of nodes belonging to the community calculated within the community \\
     \cline{2-3}
     & max\_group\_eccentricity$+\ast$ & The maximum value of eccentricity of nodes belonging to the community calculated within the community \\
     \cline{2-3}
     & sum\_network\_degree\_in & The sum of indegree of nodes belonging to the community calculated within the network \\
     \cline{2-3}
     & avg\_network\_degree\_in & The average value of indegree of nodes belonging to the community calculated within the network \\
     \cline{2-3}
     & min\_network\_degree\_in & The minimum value of indegree of nodes belonging to the community calculated within the network \\
     \cline{2-3}
     & max\_network\_degree\_in & The maximum value of indegree of nodes belonging to the community calculated within the network \\
     \cline{2-3}
     & sum\_network\_degree\_out & The sum of outdegree of nodes belonging to the community calculated within the network \\
     \cline{2-3}
     & avg\_network\_degree\_out & The average value of outdegree of nodes belonging to the community calculated within the network \\
     \cline{2-3}
     & min\_network\_degree\_out & The minimum value of outdegree of nodes belonging to the community calculated within the network \\
     \cline{2-3}
     & max\_network\_degree\_out & The maximum value of outdegree of nodes belonging to the community calculated within the network \\
     \cline{2-3}
     & sum\_network\_degree\_total$+\ast$ & The sum of total degree of nodes belonging to the community calculated within the network \\
     \cline{2-3}
     & avg\_network\_degree\_total$+\ast$ & The average value of total degree of nodes belonging to the community calculated within the network \\
     \cline{2-3}
     & min\_network\_degree\_total$+\ast$ & The minimum value of total degree of nodes belonging to the community calculated within the network \\
     \cline{2-3}
     & max\_network\_degree\_total$+\ast$ & The maximum value of total degree of nodes belonging to the community calculated within the network \\
     \cline{2-3}
     & sum\_network\_betweenness $+\ast$& The sum of betweenness of nodes belonging to the community calculated within the network \\
     \cline{2-3}
     & avg\_network\_betweenness$+\ast$ & The average value of betweenness of nodes belonging to the community calculated within the network \\
     \cline{2-3}
     & min\_network\_betweenness$+\ast$ & The minimum value of betweenness of nodes belonging to the community calculated within the network \\
     \cline{2-3}
     & max\_network\_betweenness$+\ast$ & The maximum value of betweenness of nodes belonging to the community calculated within the network \\
     \cline{2-3}
     & sum\_network\_closeness$+\ast$ & The sum of closeness of nodes belonging to the community calculated within the network \\
     \cline{2-3}
     & avg\_network\_closeness$+\ast$ & The average value of closeness of nodes belonging to the community calculated within the network \\
     \cline{2-3}
     & min\_network\_closeness$+\ast$ & The minimum value of closeness of nodes belonging to the community calculated within the network \\
     \cline{2-3}
     & max\_network\_closeness$+\ast$ & The maximum value of closeness of nodes belonging to the community calculated within the network \\
     \cline{2-3}
     & sum\_network\_eigenvector$+\ast$ & The sum of eigenvector of nodes belonging to the community calculated within the network \\
     \cline{2-3}
     & avg\_network\_eigenvector$+\ast$ & The average value of eigenvector of nodes belonging to the community calculated within the network \\
     \cline{2-3}
     & min\_network\_eigenvector$+\ast$ & The minimum value of eigenvector of nodes belonging to the community calculated within the network \\
     \cline{2-3}
     & max\_network\_eigenvector$+\ast$ & The maximum value of eigenvector of nodes belonging to the community calculated within the network \\
     \cline{2-3}
     & avg\_group\_coefficient\cite{Wasserman1994}$+\ast$ & The average of the local clustering coefficients of all the nodes in the community \\
     \cline{2-3}
     & avg\_network\_coefficient\cite{Wasserman1994}$+\ast$ & The average of the local clustering coefficients of all the nodes in the network \\
     \hline
    \multirow{38}{*}{Group/mesoscopic} & group\_size$+\ast$ & The number of nodes in the group \\
     \cline{2-3}
     & group\_density\cite{Wasserman1994}$+\ast$ & The number of connections between nodes in the group in relation to all possible connections between them \\
     \cline{2-3}
     & group\_cohesion\cite{White2001}$+\ast$ & The vertex connectivity of the community \\
     \cline{2-3}
     & group\_coefficient\_global\cite{Wasserman1994}$+\ast$ & The ratio of the triangles and the connected triples in the community \\
     \cline{2-3}
     & group\_reciprocity\cite{Newman2010} & A fraction of edges that are reciprocated within the community \\
     \cline{2-3}
     & group\_leadership\cite{Freeman1978}$+\ast$ & A measure describing centralization in the community (the largest value is for a star network) \\
     \cline{2-3}
     & neighborhood\_out & The number of nodes outside the community that have incoming connection from the nodes inside the community divided by the number of nodes in the community \\
     \cline{2-3}
     & neighborhood\_in & The number of nodes outside the community that have outgoing connection to the nodes inside the community divided by the number of nodes in the community \\
     \cline{2-3}
     & neighborhood\_all$+\ast$ & The number of nodes outside the community that are  connected to the nodes inside the community divided by the number of nodes in the community \\
     \cline{2-3}
     & group\_adhesion\cite{White2001}$+\ast$ & The minimum number of edges needed to be removed to obtain a community which is not strongly connected \\
     \cline{2-3}
     & alpha\cite{Brodka2013a} & The GED inclusion measure of group $G_i$ from time window $T_n$ in group $G_j$ from $T_{n+1}$ \\
     \cline{2-3}
     & beta\cite{Brodka2013a} & The GED inclusion measure of group $G_j$ from time window $T_{n+1}$ in group $G_i$ from $T_n$ \\
     \cline{2-3}
     & network\_ratio\_size$+\ast$ & The ratio of \emph{group\_size} to \emph{network\_size} \\
     \cline{2-3}
     & network\_ratio\_density$+\ast$ & The ratio of \emph{group\_density} to \emph{network\_density} \\
     \cline{2-3}
     & network\_ratio\_cohesion$+\ast$ & The ratio of \emph{group\_cohesion} to \emph{network\_cohesion} \\
     \cline{2-3}
     & network\_ratio\_coefficient\_global$+\ast$ & The ratio of \emph{group\_coefficient\_global} to \emph{network\_coefficient\_global} \\
     \cline{2-3}
     & network\_ratio\_coefficient\_average$+\ast$ & The ratio of \emph{group\_clustering\_coefficient} to \emph{network\_clustering\_coefficient} \\
     \cline{2-3}
     & network\_ratio\_reciprocity & The ratio of \emph{group\_reciprocity} to \emph{network\_reciprocity} \\
     \cline{2-3}
     & network\_ratio\_leadership$+\ast$ & The ratio of \emph{group\_leadership} to \emph{network\_leadership} \\
     \cline{2-3}
     & network\_ratio\_eccentricity$+\ast$ & The ratio of \emph{avg\_group\_eccentricity} to \emph{network\_avg\_eccentricity} \\
     \cline{2-3}
     & network\_ratio\_adhesion$+\ast$ & The ratio of \emph{group\_adhesion} to \emph{network\_adhesion} \\
     \cline{2-3}
     & \textbf{phys\_rev}$\ast$ & \textbf{The number of articles belonging to the group that were published in the Physical Review journal} \\
     \cline{2-3}
     & \textbf{phys\_rev\_a}$+\ast$ & \textbf{The number of articles belonging to the group that were published in the Physical Review A journal} \\
     \cline{2-3}
     & \textbf{phys\_rev\_b}$+\ast$ & \textbf{The number of articles belonging to the group that were published in the Physical Review B journal} \\
     \cline{2-3}
     & \textbf{phys\_rev\_c}$+\ast$ & \textbf{The number of articles belonging to the group that were published in the Physical Review C journal} \\
     \cline{2-3}
     & \textbf{phys\_rev\_d}$+\ast$ & \textbf{The number of articles belonging to the group that were published in the Physical Review D journal} \\
     \cline{2-3}
     & \textbf{phys\_rev\_e}$+\ast$ & \textbf{The number of articles belonging to the group that were published in the Physical Review E journal} \\
     \cline{2-3}
     & \textbf{phys\_rev\_lett}$+\ast$ & \textbf{The number of articles belonging to the group that were published in the Physical Review Letters journal} \\
     \cline{2-3}
     & \textbf{phys\_rev\_stab}$+\ast$ & \textbf{The number of articles belonging to the group that were published in the Physical Review STAB journal} \\
     \cline{2-3}
     & \textbf{phys\_rev\_stper}$+$ & \textbf{The number of articles belonging to the group that were published in the Physical Review STPER journal} \\
     \cline{2-3}
     & \textbf{physics}$\ast$ & \textbf{The number of articles belonging to the group that were published in the Physics journal} \\
     \cline{2-3}
     & \textbf{rev\_mod\_phys}$+\ast$ & \textbf{The number of articles belonging to the group that were published in the Review of Modern Physics journal} \\
     \cline{2-3}
     & \textbf{sum\_group\_age}$+\ast$ & \textbf{The sum of age of articles belonging to the group. In the co-reference network the age of an article is the average age of the articles it references to. In the co-citation network the age of an article is the age of the articles being cited.} \\
     \cline{2-3}
     & \textbf{avg\_group\_age}$+\ast$ & \textbf{The average age of articles belonging to the group} \\
     \cline{2-3}
     & \textbf{min\_group\_age}$+\ast$ & \textbf{The minimum age of articles belonging to the group} \\
     \cline{2-3}
     & \textbf{max\_group\_age}$+\ast$ & \textbf{The maximum age of articles belonging to the group} \\
     \cline{2-3}
     & \textbf{network\_ratio\_avg\_group\_age}$+\ast$ & \textbf{The ratio of avg\_group\_age to the average age of all articles in the network} \\
     \cline{2-3}
     & time\_window$+\ast$ & The number of time window from which the community instance was obtained \\
     \hline
    \multirow{9}{*}{Network/macroscopic} & network\_size$+\ast$ & The number of nodes in the network \\
    \cline{2-3}
     & network\_density$+\ast$ & The number of connections between nodes in the network in relation to all possible connections between them \\
     \cline{2-3}
     & network\_cohesion$+\ast$ & The vertex connectivity of the network \\
     \cline{2-3}
     & network\_coefficient\_global$+\ast$ & The ratio of the triangles and the connected triples in the network \\
     \cline{2-3}
     & network\_coefficient\_average$+\ast$ & The average of the local clustering coefficients of all the nodes in the network \\
     \cline{2-3}
     & network\_reciprocity & A fraction of edges that are reciprocated within the network \\
     \cline{2-3}
     & network\_leadership$+\ast$ & A measure describing centralization in the network (the largest value is for a star network) \\
     \cline{2-3}
     & network\_avg\_eccentricity$+\ast$ & The average value of eccentricity of nodes within the network. \\
     \cline{2-3}
     & network\_adhesion$+\ast$ & The minimum number of edges needed to be removed to obtain a graph which is not strongly connected \\
    \hline
  \caption{List of all features used in the study. Features proposed in this study are shown in bold.}\label{tab:TabS1}
\end{longtable}
\end{center}

\end{document}